\documentstyle[12pt,a4,epsf,bm]{article}

\def\SU{\mathop{\rm SU}}
\def\U{\mathop{\rm U}}
\def\SL{\mathop{\rm SL}}
\def\sign{\mathop{\rm sign}\nolimits}
\def\tr{\mathop{\rm tr}\nolimits}
\def\mod{\mathop{\rm mod}\nolimits}

\newcommand{\ol}{\overline}
\newcommand{\wh}{\widehat}
\newcommand{\wt}{\widetilde}

\begin{document}

\begin{titlepage}
\title{\hfill\parbox{4cm}
       {\normalsize UT-06-21\\{\tt hep-th/0609163}\\Sep 2006}\\
       \vspace{3cm}
       Global symmetries and 't~Hooft anomalies in brane tilings
       \vspace{2cm}}
\author{Yosuke Imamura\thanks{E-mail: \tt imamura@hep-th.phys.s.u-tokyo.ac.jp}%
\\[20pt]
{\it Department of Physics, University of Tokyo, Tokyo 113-0033, Japan}
}
\date{}

\maketitle
\thispagestyle{empty}

\vspace{0cm}

\begin{abstract}
\normalsize
We investigate the relation between gauge theories and brane configurations
described by brane tilings.
We identify $\U(1)_{\rm B}$ (baryonic), $\U(1)_{\rm M}$ (mesonic), and $\U(1)_{\rm R}$
global symmetries in gauge theories with gauge symmetries
in the brane configurations.
We also show that $\U(1)_{\rm M}\U(1)_{\rm B}^2$ and $\U(1)_{\rm R}\U(1)_{\rm B}^2$ 't~Hooft anomalies
are reproduced as gauge transformations of the classical brane action.
\end{abstract}

\end{titlepage}

\section{Introduction}
String theory is a useful tool to study dynamics of gauge theories.
An advantage of using string theory for the analysis of
gauge theories is
that we can translate
quantum corrections and non-perturbative effects
in gauge theories to classical phenomena associated with
geometries of spacetime and brane configurations.

Among many kinds of dualities,
the duality between conifolds and ${\cal N}=1$ superconformal gauge theories
attracts a great deal of attention.
The first non-trivial example of this duality was proposed
in \cite{Klebanov:1998hh}.
It is the duality between the conifold over $T^{1,1}$ and
a certain $\SU(N)\times\SU(N)$ superconformal gauge theory
at the IR fixed point.
This duality has been generalized to more complicated ones.
The recent discovery of the explicit metrics of classes
of Sasaki-Einstein manifolds\cite{Gauntlett:2004yd,Cvetic:2005ft,Martelli:2005wy}
provides us many examples
of dualities we can explicitly check the validity.
For example, it has been confirmed that
the volumes of the Sasaki-Einstein manifolds
and some supersymmetric cycles in them
correctly reproduce the
central charges and conformal dimensions of baryonic operators
in superconformal gauge theories, which can be determined
on the field theory side with the help of the $a$-maximization
technique\cite{Intriligator:2003jj}.

There are also many attempts to generalize the correspondence
to non-conformal cases.
One way to break the conformal symmetry on the field theory side
is to change the ranks
of the $\SU(N)$ gauge groups.
This is realized by the introduction of fractional D3-branes
on the gravity side.
The
coupling running\cite{Klebanov:1999rd,Klebanov:2000nc}
and the duality cascade\cite{Klebanov:2000hb}
caused by the introduction of the fractional branes are studied
by using dual gravity solutions.

The brane tilings\cite{Hanany:2005ve,Franco:2005rj,Franco:2005sm,Hanany:2005ss,Feng:2005gw,Franco:2006gc}, proposed by Hanany et al. are different,
but closely related way
to realize ${\cal N}=1$ quiver gauge theories
with fivebranes in type IIB theory.
The brane configurations are conveniently described with tilings on tori.
Each face is identified with a stack of $N$ D5-branes,
and an $\SU(N)$ gauge group lives on it.
An edge shared by two faces
represents the intersection of the D5-branes and NS5-branes, and corresponds to
a chiral multiplet belonging to a bi-fundamental representation
of the two gauge groups associated with the two faces.
Thus, the tilings can be regarded as dual graphs of quiver diagrams
drawn on tori, which are often called periodic quiver diagrams.
An advantage of the brane tilings (and the periodic quiver diagrams)
to the ordinary quiver diagrams
is that we can easily read off the superpotential
from the diagram.

There are many rules proposed to read off the properties
and phenomena in gauge theories
from the diagrams.
For example, we can easily obtain anomaly free charge assignments
of global symmetries from the brane tilings\cite{Butti:2005vn}.
We can also determine IR behavior of non-conformal
gauge theories depending on the rank distributions\cite{Butti:2006hc}.
It is important to obtain these relations between
the graphical information in the tilings and
the properties of gauge theories
directly by using action or equations of motion of
branes and supergravity.
In this paper, we particularly discuss how anomalies in gauge theories
show up in the brane tilings.
Anomalies arise as one-loop corrections in gauge theories,
and have topological nature in the sense that
their coefficients are quantized, and are invariant under
continuous variation of parameters.
This makes analysis on the string side tractable.
In general it is difficult to determine the precise shape of
branes in a brane system.
By the reason we mention above, for the analysis of anomalies,
we do not need the precise shape of the branes.
We only need the topological structure and
the asymptotic shape of branes.
In \cite{Imamura:2006ub}, the relation between cancellations of
gauge ($\SU(N_c)^3$) and $\U(1)_B\SU(N_c)^2$ anomalies and flux conservations
on the brane system is discussed, and it is shown that
if the boundary conditions imposed on the fields on branes at
fivebrane junctions are satisfied, the anomaly cancellations are automatically
realized.
In this paper, we discuss
so-called 't~Hooft anomalies which are in general not canceled.
We show that some of the anomalies can be reproduced
as the variations of the classical brane action by gauge transformations.
(In this paper we are not very careful about signs.)

This paper is organized as follows.
In \S\ref{config.sec}, we briefly explain the relation
between the tiling diagrams and the structure of
the branes in the system.
In \S\ref{global.sec}, we identify global $\U(1)$
symmetries in gauge theories
with gauge symmetries in the brane system.
Using this identification, we compute some of 't~Hooft anomalies,
in \S\ref{anomaly.sec}.
We only discuss $\U(1)_B^3$, $\U(1)_M\U(1)_B^2$ and $\U(1)_R\U(1)_B^2$
anomalies, where $\U(1)_B$, $\U(1)_M$, and $\U(1)_R$ are the
baryonic, mesonic (flavor), and R-symmetries, respectively,
which are defined in \S\ref{global.sec}.
Among these three classes of anomalies, the $\U(1)_B^3$ anomalies are known to vanish,
and we use this fact to fix the ambiguity of the regularization.
We show that the other two kinds of anomalies are reproduced
as variations of the classical brane action.
The last section is devoted for discussions.
In Appendix \ref{junc.sec}, we discuss the action of fivebrane junctions
and boundary conditions imposed on the gauge fields.
In Appendix \ref{maxwell.sec}, we solve a differential equation
which appears in \S\ref{anomaly.sec}.

\section{Brane tilings and D5-NS5 systems}\label{config.sec}
\begin{figure}[htb]
\epsfxsize=8cm
\centerline{\epsfbox{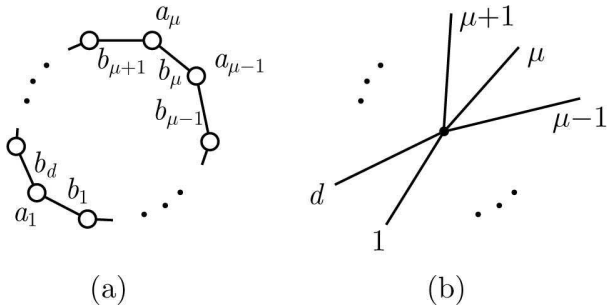}}
\caption{A toric diagram (a) and the corresponding web diagram (b).}
\label{mu.eps}
\end{figure}

The toric diagrams and web diagrams (Figure \ref{mu.eps}) are often used to describe structure of
toric Calabi-Yau manifolds, and we can obtain information of the
corresponding quiver gauge theories from these diagrams.
Toric diagrams are convex polygons in a two-dimensional lattice.
Let $d$ be the number of edges on the perimeter of a toric diagram.
We label these $d$ edges with $\mu,\nu=1,\ldots,d$
in counter clockwise order.
We also use the label $\mu$ for the vertex between the edges $\mu$ and $\mu+1$.
(Figure \ref{mu.eps})
The indices are defined modulo $d$,
and $\mu=d+1$ is identified with $\mu=1$.

The web diagrams are graphical dual to the toric diagrams.
Each external leg in a web diagram corresponds to an edge of the dual toric diagram,
and we label the legs with $\mu$ like the edges of the
toric diagram.
In general, web diagrams may have internal lines and loops.
In this paper we are interested only in the external legs,
and web diagrams are represented as sets of semi-infinite radial lines.

By taking T-duality transformation along specific two cycles,
a toric Calabi-Yau geometry is transformed into an NS5-brane system,
and D3-branes at the tip of the cone
are mapped to D5-branes wrapped on ${\bf T}^2$.
Let $x^M$ ($M=0,\ldots,9$) be the coordinates
of $10$-dimensional spacetime.
The $4$-dimensional gauge theory is defined in ${\bf R}^4$
along $0123$.
The directions $x^5$ and $x^7$ are compactified
and the T-duality is taken along these directions.
The four dimensional space along $4567$ is
topologically $({\bf C}^\times)^2$. (Table \ref{brane.tbl})

The D5-brane world volume is ${\bf R}^4\times T$
where $T$ is the torus along 57 directions,
and the NS5-brane world volume is ${\bf R}^4\times\Sigma$,
where $\Sigma$ is a $2$-dimensional surface
in the $4567$ space.
\begin{table}[htb]
\caption{The brane configuration corresponding to brane tilings.
$5$ and $7$ are compactified.
$\Sigma$ is a two dimensional surface in the 4567 space.}
\label{brane.tbl}
$$
\begin{array}{ccccc|cccc|cc}
\hline
\hline
           & 0 & 1 & 2 & 3 & 4 & 5 & 6 & 7 & 8 & 9 \\
\hline
\mbox{D5} & \circ & \circ & \circ & \circ & & \circ & & \circ \\
\mbox{NS5} & \circ & \circ & \circ & \circ & \multicolumn{4}{c|}{\Sigma} \\
\hline
\end{array}
$$
\end{table}
Because any branes we consider here always spread along $0123$,
we mainly focus only on the internal part of the worldvolumes,
$T$ and $\Sigma$,
and call these two-dimensional surfaces simply worldvolumes.

In the weak coupling limit, $g_{\rm str}\rightarrow0$,
in which the NS5-brane tension is much
larger than the D5-brane tension,
the NS5-brane world volume $\Sigma$
is a holomorphic curve in the $({\bf C}^\times)^2$
described by the Newton polynomial associated with
the toric diagram.
The projection of the surface to the non-compact $46$-plane is called amoeba.
It is easily shown that $\Sigma$ generically has $d$ punctures, and
they are represented as infinitely long thorns of amoeba.
The web diagram can be regarded as the ``tropicalization'' of the amoeba
in which the thorns becomes semi-infinite radial lines.
These lines
are semi-infinite cylinders of NS5-branes,
and
the surface $\Sigma$ can be constructed as the union
of these $d$ semi-infinite cylinders.
We refer to these cylinders as faces in $\Sigma$.
They are topologically punctured disks, and are labeled by
$\mu=1,\ldots,d$
in the same way as the external legs in the web diagram.

\begin{figure}[htb]
\epsfxsize=8cm
\centerline{\epsfbox{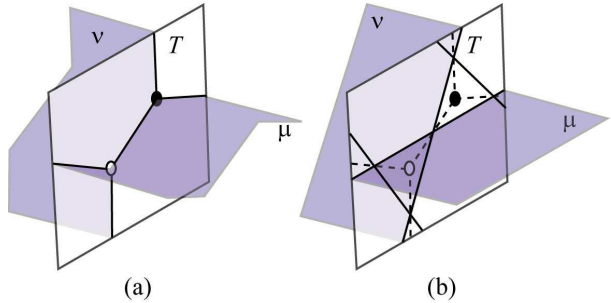}}
\caption{(a) Structure of branes around an edge. This is correct only topologically. (b) The real shape of the brane configuration in the strong coupling limit.}
\label{zigzag.eps}
\end{figure}

In order to determine the real shapes of both the NS5 and D5 branes,
we need to solve
the equations of motion of the branes (or BPS equations equivalent to them)
and in general it is not easy.
In some cases, we are interested only in the topological%
\footnote{In this paper, we use the term ``topological'' in the
following sense:
If two brane configurations are topologically the same,
we can continuously deform one to the other.
In general, these two may have different geometrical topology.}
structure of branes.
In such cases, instead of considering the real shape of branes,
it is convenient to consider another configuration obtained
from the real shape by continuous deformation
which does not change the asymptotic shape
of the system.
In this paper we mainly use configurations in which the D5-brane world volume $T$
is the flat torus, and the intersection
$T\cap\Sigma$ is a bipartite graph ((a) in Fig. \ref{zigzag.eps}).
A bipartite graph is a graph in which
all the vertices can be colored with two colors, say, black and white,
in such a way that any two vertices
connected by an edge have different colors.
The bipartite graph drawn on $T$ is called ``brane tilings''.

\begin{figure}[htb]
\epsfxsize=5cm
\centerline{\epsfbox{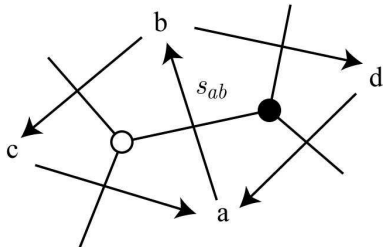}}
\caption{The oriented segments (arrows) corresponding to the bi-fundamental fields}
\label{sab.eps}
\end{figure}

Given a brane tiling, we can easily read off the information of
the corresponding quiver gauge theory.
The faces in the tiling, each of which is a stack of
$N$ D5-brane disks,
represent the $\SU(N)$ factors in the gauge group.
We use $a,b,\ldots$ for labeling the faces in the tiling,
and denote the gauge group associated with the face $a$ by $\SU(N_a)$.
Edges correspond to bi-fundamental chiral multiplets.
Let $I$ be the edge shared by two faces $a$ and $b$.
The chiral multiplet $\Phi_I$ corresponding to the edge belongs
to a bi-fundamental representation of $\SU(N_a)\times\SU(N_b)$.
This arises as massless modes of open strings stretched between
D5-branes on faces $a$ and $b$\cite{Elitzur:2000pq}.
These open strings graphically represented as an oriented segment
connecting faces $a$ and $b$, and
we denote the segment by $s_{ab}$.
(Fig. \ref{sab.eps})
These segments are nothing but the arrows in the corresponding
periodic quiver diagram,
and if the orientation is from $b$ to $a$ (from $a$ to $b$),
the chiral multiplet belongs to
$({\bf N}_a,\ol{\bf N}_b)$
($(\ol{\bf N}_a,{\bf N}_b)$).

The orientations of the segments are determined according to the
colors of vertices.
We take clockwise orientation for arrows around a black vertex,
and counter clockwise orientation for arrows around a white vertex.
Because of the bipartiteness of the graph we can
consistently and uniquely determine the orientation of the arrows by
this rule.
To represent this orientation, we define the function
$\sign(a,b)$ for a pair of faces $a$ and $b$ sharing an edge
in such a way that
if the orientation is from $b$ to $a$ (from $a$ to $b$)
the function is $\sign(a,b)=+1$ ($-1$).
In general, two faces may share more than one edge.
In such a case, we should define $\sign(a,b)$ for each edge separately.
Although $\sign(a,b)$ depends not only on $a$ and $b$
but also on the edges, we will not give it explicitly as argument.

\begin{figure}[htb]
\epsfxsize=5cm
\centerline{\epsfbox{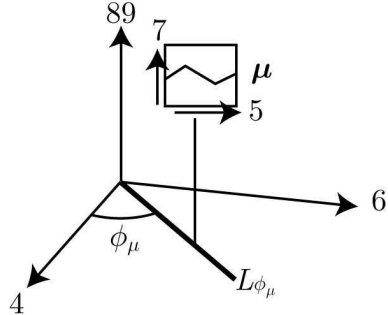}}
\caption{Each leg in the web diagram represents an NS5-brane.
It is semi-infinite radial line $L_{\phi_\mu}$ in the non-compact 4689 space.
In the internal space along 57, it is wrapped on the cycle ${\bm\mu}$.}
\label{lphi2.eps}
\end{figure}

The boundaries of faces on $\Sigma$ are
zig-zag paths in the bipartite graph on T.
((a) in Figure \ref{zigzag.eps})
A zig-zag path is a closed oriented path consisting of edges
in a bipartite graph drawn on
an orientable surface
which turns most left at black vertices and turns most right at white vertices.
Let $\bm\mu$ represent the zig-zag path on $T$
which is the boundary of a face $\mu$ in $\Sigma$.
The path ${\bm\mu}$ belongs to a non-trivial homology class
of $T$,
and is identified with
the vector $v_\mu-v_{\mu-1}$,
where $v_\mu$ is the two-dimensional integral coordinate vector for
the vertex $\mu$ in the toric diagram.
We can represent the faces in $\Sigma$,
the semi-infinite cylinders of NS5-branes,
as the direct product of the zig-zag paths
$\bm\mu$ in the torus
and the semi-infinite radial lines $L_{\phi_\mu}$ in the 46 plane, where $\phi_\mu$ is the direction of the external line $\mu$.
(Figure \ref{lphi2.eps})
The real shape of the worldvolumes of NS5-branes are of course smooth
and their sections are never zig-zag lines.
We here, however, are interested in the topological structure,
and do not distinguish between them.

An edge in a tiling is always shared by two zig-zag paths in $T$.
In this paper, we assume that if a pair of two zig-zag paths
have more than one intersection
they intersect in the same orientation at all the intersections. 
It follows this assumption that
any zig-zag path does not have self-intersections.
This is necessary for the graph to give a consistent quiver gauge theory\cite{Hanany:2005ss}.
The number of edges shared by two zig-zag paths $\bm\mu$ and $\bm\nu$
is $|\langle{\bm\mu},{\bm\nu}\rangle|$
where $\langle{\bm\mu},{\bm\nu}\rangle$ is the
intersection number of the two paths.
If the cohomology classes for the paths $\bm\mu$ and $\bm\nu$
are given as the linear combination of basis
$({\bm\alpha},{\bm\beta})$ by
\begin{equation}
{\bm\mu}=p_1{\bm\alpha}+q_1{\bm\beta},\quad
{\bm\nu}=p_2{\bm\alpha}+q_2{\bm\beta},
\end{equation}
the intersection number is given by
\begin{equation}
\langle{\bm\mu},{\bm\nu}\rangle=p_1q_2-p_2q_1.
\end{equation}
(We use $\bm\mu$ and $\bm\nu$ for two meanings,
zig-zag paths and homology classes for the paths.)
Because every edge is shared by two zig-zag paths,
the total number of
the bi-fundamental matter fields is given by
\begin{equation}
N_{\rm matter}=\frac{1}{2}\sum_{\mu,\nu}|\langle{\bm\mu},{\bm\nu}\rangle|.
\label{nmatter}
\end{equation}
Similarly to the function $\sign(a,b)$,
We define the signature function for a pair of two faces $\mu$ and $\nu$ on $\Sigma$
by
$\sign(\mu,\nu)\equiv\sign(\langle{\bm\mu},{\bm\nu}\rangle)$.

Because the bipartite graph is the intersection of
$T$ and $\Sigma$, we can also regard it as a graph on $\Sigma$.
This gives another set of zig-zag paths
because the definition of zig-zag paths depends on
the choice of the orientable surface on which the graph is drawn.
We can easily see that zig-zag paths
defined in $\Sigma$ are boundaries of faces on $T$.
Let $\bm a$ be the zig-zag path on $\Sigma$ corresponding to the
boundary of face $a$.

Because $\Sigma$ is an orientable surface as well as $T$,
we can define ${\bf Z}$-valued
intersection number for two zig-zag paths $\bm a$ and $\bm b$.
We denote it by $\langle{\bm a},{\bm b}\rangle$.
The signature $\sign(a,b)$ we defined above for a pair of faces $a$ and $b$
on $T$ is identical with $\sign(\langle{\bm a},{\bm b}\rangle)$.

For the zig-zag paths on $T$, the following relation holds:
\begin{equation}
\sum_\mu{\bm\mu}=0\quad
\mbox{(as a homology class of $T$).}
\label{totalmu}
\end{equation}
This is because all the edges are shared by two zig-zag paths
and these two paths have opposite orientation
on the edge.
A similar relation holds for the zig-zag paths on $\Sigma$.
The zig-zag path on $\Sigma$ satisfy
\begin{equation}
\sum_a{\bm a}=0\quad
\mbox{(as a homology class of $\ol\Sigma$),}
\end{equation}
where $\ol\Sigma$ is the closure of $\Sigma$.
This is because the boundaries of adjacent faces have opposite
orientation along the shared edge.

We should emphasize that the structure of branes we describe above
is correct only in the topological sense.
The real structure would be difficult to obtain in general.
The D5 and NS5 worldvolumes are deformed by the effect of the other branes,
and become two surfaces in $({\bf C}^\times)^2$ sharing part of them.
In some cases, however, we can easily determine the shape of
branes.
One is the weak coupling limit we mentioned above.
The opposite limit is also interesting.
In the strong coupling limit, in which the D5-brane tension is
much larger than the NS5-brane tension,
the system consists of almost flat branes.
It looks like (b) in Figure \ref{zigzag.eps}.
In addition to the original faces of the bipartite graph,
new faces are formed around the vertices.
The white and black vertices become polygons
of $(N,1)$ and $(N,-1)$ fivebrane, respectively.

Because the edges in brane tilings are the intersection
of the NS5-brane $\Sigma$ and the stack of $N$ D5-branes $T$,
each edge can be regarded as a $4$-junction of $5$-branes.
(We here treat a stack of $N$ D5-branes as a single $5$-brane with the D5-brane
charge $N$.)
The 5-brane charges of four $5$-branes meeting
at the $4$-junction
are $(0,1)$, $(0,1)$, $(N,0)$ and $(N,0)$.
This can be generalized into more general junctions.
For example, we can change the fivebrane charges at a $4$-junction
to
\begin{equation}
(N_a,0),\quad
(N_b,0),\quad
(p_\mu,1),\quad
(p_\nu,1),
\end{equation}
where $a$ and $b$ are faces on $T$ sharing an edge
and $\mu$ and $\nu$ are faces in $\Sigma$
sharing the same edge.
The D5-brane charges $N_a$, $N_b$, $p_\mu$ and $p_\nu$ must satisfy
the charge conservation condition
\begin{equation}
\sign(a,b)(N_a-N_b)+\sign(\mu,\nu)(p_\mu-p_\nu)=0.
\label{nprel}
\end{equation}
By this generalization, the numbers of D5-branes depend on faces.
This corresponds to the introduction of
fractional D3-branes in the dual Calabi-Yau cone.
On the gauge theory side, this gives gauge groups
with different ranks.

The boundary condition (\ref{nprel})
guarantees the D5-brane charge conservation at junctions,
and we can derive constraints imposed on $N_a$ and $p_\mu$.
The D5 charge conservation on $T$
gives
\begin{equation}
\sum_\mu p_\mu{\bm\mu}=0\quad
\mbox{(as a homology class of $T$).}
\label{pmrel}
\end{equation}
For a given set of $\bm\mu$,
this imposes two independent conditions on the set of numbers $p_\mu$,
and the number of independent components in $p_\mu$ is $d-2$.
This condition is useful when we classify the fractional branes.
Deformations of the complex structure of Calabi-Yau manifold is
known to be described by Altmann's rule,
and it corresponds to splittings of NS5-brane system.
If a set of integers $p_\mu$ satisfies the condition (\ref{pmrel})
in each component of the system, the corresponding fractional brane
is called deformation fractional brane\cite{Butti:2006hc}.

We also have the following constraint imposed on $N_a$ from the cancellation of
D5-charge flowing into $\Sigma$.
\begin{equation}
\sum_a N_a{\bm a}=0\quad
\mbox{(as a homology class of $\ol\Sigma$).}
\label{narel}
\end{equation}
A set of $N_a$ satisfying the relation
(\ref{narel})
gives
an anomaly-free rank distribution\cite{Imamura:2006ub}.
If there are $\SU(2)$ factors in the gauge group,
this guarantees the absence of the global anomaly\cite{Witten}.

In the following sections, we restrict our attention to the
conformal case with all ranks the same.

\section{Global symmetries}\label{global.sec}
The global $\U(1)$ symmetries of ${\cal N}=1$ quiver gauge theories
are classified in three classes according to operators
they non-trivially act on.
A global symmetry rotating the supercharge is called R-symmetry,
and denoted  by $\U(1)_R$.
Global symmetries which do not act on the supercharge
are called flavor symmetries,
and they are divided into two groups,
mesonic and baryonic symmetries.%
\footnote{Flavor symmetries often mean what
are referred to as mesonic symmetries
in this paper.
We here use this term to represent
baryonic and mesonic symmetries.}

These global symmetries should be realized as gauge symmetries
in string theory.
The purpose of this section is to identify the gauge symmetries
corresponding to the global symmetries in gauge theories.
Because global symmetries are specified by charge assignments to matter fields,
we first discuss charged objects in the brane systems
corresponding to matter fields.

For later use we define some ``delta functions''.
$\delta(a)$ is the function in $T$ which is $1$ in the face $a$, and
vanishes in the other faces.
$\delta(\mu)$ is the function in $\Sigma$ similarly defined.
$\delta(I)$ is the one-form delta function
on $T$
supported by the edge $I$.
The signature is chosen so that $\int_{s_{ab}}\delta(I)=1$
where $a$ and $b$ are the two faces sharing $I$
and $s_{ab}$ is the oriented segment defined in \S\ref{config.sec}.
$\delta({\bm\mu})$ is the closed $1$-form delta function
on $T$
with support on $\bm\mu$.
The integral of $\delta({\bm\mu})$ along a path $C$
gives the intersection of $C$ and $\bm\mu$.
The following relation holds:
\begin{equation}
\int_T\delta({\bm\mu})\wedge\delta({\bm\nu})
=\int_{\bm\mu}\delta({\bm\nu})
=\langle{\bm\mu},{\bm\nu}\rangle.
\label{deltaint}
\end{equation}
$\delta({\bm a})$ is the closed $1$-form delta function
on $\Sigma$ with support on a zig-zag path $\bm a$.
This also satisfies the similar relation to (\ref{deltaint}).

\subsection{Strings and chiral operators}\label{strandop.ssec}
Let $Q_I$ be a charge assignment of a flavor symmetry
to fields $\Phi_I$.
We define the one-form $\cal Q$ on the torus $T$ by
\begin{equation}
{\cal Q}\equiv\sum_IQ_I\delta(I).
\end{equation}
Each term in the superpotential corresponds to the vertices
in the tiling.
The term corresponding to a vertex is trace of product of
bi-fundamental field associated with edges
around the vertex.
For the superpotential to be invariant under the symmetry
specified by the charge assignment,
the sum of charges for edges sharing one vertex must vanish.
Therefore, we need to require the one form satisfy
\begin{equation}
d{\cal Q}=0.
\label{closeQ}
\end{equation}

There are two kinds of gauge invariant chiral operators
made of bi-fundamental chiral multiplets.
Operators in one kind are called mesonic operators.
These are the trace of the products of bi-fundamental fields.
Each mesonic operator can naturally be associated with
the closed path made of the oriented segments
corresponding to the constituent bi-fundamental fields.
Because each segment represents open string,
it is possible to regard the closed paths as closed strings.
Even though the process in which open strings merge into
an closed string is suppressed in the decoupling limit,
the identification of closed strings and mesonic operators is convenient
because
the closed strings carry the same charge with the mesonic operators.

Let $P_{\cal O}$ be the closed path corresponding to a mesonic operator
${\cal O}$.
The $\U(1)$ charge of the operator ${\cal O}$
can be obtained as the contour integral
of the one-form $\cal Q$ along the path $P_{\cal O}$ corresponding to the operator:
\begin{equation}
Q({\cal O})=\oint_{P_{\cal O}}{\cal Q}.
\label{mesoncharge}
\end{equation}
Because of the condition (\ref{closeQ}),
the charge of mesonic operator depends only on the homology class
of the corresponding contour.
This means that if two one-forms ${\cal Q}$
for two flavor symmetries belong to the same
cohomology class we cannot distinguish between them
by using couplings to mesonic operators.
In order to distinguish between such two symmetries
we need to use
baryonic operators.

The baryonic operators are operators constructed by using
determinant with respect to the color indices.
We can identify the baryonic operator $\det\Phi_I$ with
D-strings stretched between two faces $\mu$ and $\nu$ on $\Sigma$
sharing the edge $I$.
One way to confirm this is to consider the T-duality to the Calabi-Yau cone.
In the context of AdS/CFT, baryonic operators are identified with
D3-branes wrapped on $3$-cycles in Calabi-Yau\cite{Gubser:1998fp}.
Through the T-duality transformation, the wrapped $3$-branes
are mapped to D-strings ending on NS5-branes.

\begin{figure}[htb]
\epsfxsize=12cm
\centerline{\epsfbox{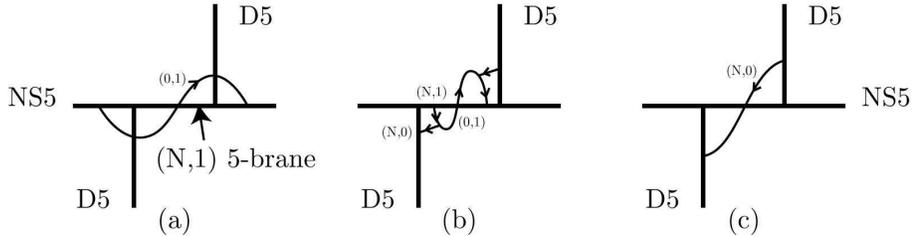}}
\caption{Transition between one D-string and $N$ F-strings.
The pairs of small numbers are string charges.}
\label{hw.eps}
\end{figure}

Another way to confirm the relation between D-strings and the
baryonic operators is to show the existence of the process in which
an open D-strings decays into $N$ open fundamental strings.
Let us consider a D-string stretched between two faces on $\Sigma$
separated by the intersection with a stack of $N$ D5-branes.
((a) in Figure \ref{hw.eps})
In the figure, we deform the four-junction of $5$-branes
into two three-junctions on which NS5, D5, and $(N,1)$ $5$-brane meet.
Because only $(N,1)$-strings can end on the $(N,1)$ $5$-brane,
if one move the endpoints of the D-string to the middle part of
horizontal line, which represents the $(N,1)$ $5$-brane,
$N$ fundamental strings are created at the both ends of the D-string
by the Hanany-Witten effect\cite{Hanany:1996ie}
so that the charge of the string endpoints on the horizontal line
become $(N,1)$.
((b) in Figure \ref{hw.eps})
After pair annihilation of the two end points on the $(N,1)$ fivebrane,
we are left with $N$ fundamental strings stretched between
two D5-branes.
((c) in Figure \ref{hw.eps})
The existence of this process
means that the D-string is a bound state of $N$ fundamental strings.

As we mentioned above, the flavor symmetries are divided in two classes.
If the one-form
${\cal Q}$ for a symmetry is in the trivial cohomology class,
mesonic operators are neutral with respect to this symmetry
and
the symmetry is called baryonic symmetry.
Otherwise, the symmetry is called mesonic symmetry.

\subsection{Baryonic symmetries}
There is a simple way to obtain
charge assignment of anomaly free flavor symmetries
with the help of the toric diagram\cite{Butti:2005vn}.
The prescription is as follows:
\begin{itemize}
\item
Associate numbers $a_\mu$
satisfying
\begin{equation}
\sum_{\mu=1}^d a_\mu=0
\label{condona1}
\end{equation}
to the vertices on the perimeter of the
toric diagram.

\item
The $\U(1)$ charge $Q_I$ of the chiral multiplet
associated with the edge $I$
is given by
\begin{equation}
Q_I=\sign(\mu,\nu)\sum_{\rho=\nu}^{\mu-1}a_\rho,
\end{equation}
where $\bm\mu$ and $\bm\nu$ are zig-zag path sharing the edge $I$.
The indices $\rho$ runs from $\nu$ to $\mu-1$ in the counter-clockwise
direction on the perimeter of the toric diagram.
\end{itemize}
As is pointed out in \cite{Butti:2006hc},
it is convenient to define parameter $b_\mu$
by
\begin{equation}
b_{\mu+1}-b_\mu=Na_\mu.
\label{bba}
\end{equation}
(The normalization of the parameters
$b_\mu$ adopted here is different from that in
\cite{Butti:2006hc} by the factor $N$.)
These parameters are associated with
the external legs of the web diagram, or equivalently
edges of the toric diagram.
Due to the condition
(\ref{condona1}) we can define $b_\mu$ satisfying this relation,
and $a_\mu$ given by (\ref{bba}) automatically satisfy
the condition (\ref{condona1}).
In terms of the parameters $b_\mu$,
the charge $Q_I$ is given by
\begin{equation}
Q_I=\frac{1}{N}\sign(\mu,\nu)(b_\mu-b_\nu).
\label{qbb}
\end{equation}

The above prescription can be used for both baryonic and mesonic symmetries.
If we want to obtain baryonic charge assignments,
we should impose
\begin{equation}
\sum_{\mu=1}^d v_\mu a_\mu=0,
\label{condona2}
\end{equation}
on the parameters $a_\mu$
in addition to (\ref{condona1}),
where $v_\mu$ is the two dimensional coordinate vector for
the vertex $\mu$ in the toric diagram.
Using ${\bm\mu}=v_\mu-v_{\mu-1}$,
the condition (\ref{condona2}) is rewritten as
\begin{equation}
\sum_\mu b_\mu{\bm\mu}=0.
\label{vcondition}
\end{equation}
These rules for obtaining mesonic and baryonic symmetries
are naturally reproduced by identifying the
corresponding gauge fields in the brane configuration.

Let us first discuss baryonic symmetries in this subsection.
This kind of symmetries do not couple to mesonic operators,
and the contour integral (\ref{mesoncharge}) vanishes.
This means that the one-form ${\cal Q}$ is exact,
and belongs to the trivial cohomology class.
Due to this,
the $1$-form can be given as
\begin{equation}
{\cal Q}=d{\cal S}
\label{qds}
\end{equation}
with a function ${\cal S}$ on the tiling
defined by
\begin{equation}
{\cal S}=\sum_aS_a\delta(a)
\end{equation}
with some number assignment $S_a$ to faces.
The relation (\ref{qds}) are
equivalent to the relation
\begin{equation}
Q_I=\sign(a,b)(S_a-S_b).
\label{qbys}
\end{equation}
The baryonic symmetries with charge assignment $Q_I$
given by (\ref{qbys})
can be realized
with the gauge field $\wh V^{\rm D5}$ on the D5-branes
given by
\begin{equation}
\wh V^{\rm D5}={\cal S}V{\bf 1}_N
\label{vd5baryon}
\end{equation}
where ${\bf 1}_N$ is the $N\times N$ unit matrix for the color indices.
We use hats to emphasize that fields are $N\times N$ matrices.
The end points of open strings on face $a$ couples to the non-dynamical
gauge field
$V$ with the charge $\pm S_a$, and the charge of open string stretched on
$s_{ab}$ is given by (\ref{qbys}).

We can regard edges as four-junctions of fivebranes
and the boundary condition imposed on gauge fields on
the four fivebranes is%
\footnote{This boundary condition is obtained from
the condition for three-junctions discussed in Appendix \ref{junc.sec}.}
\begin{equation}
\sign(a,b)\tr(\wh V_a^{\rm D5}-\wh V_b^{\rm D5})
+\sign(\mu,\nu)(V_\mu^{\rm NS5}-V_\nu^{\rm NS5})=0.
\label{abmnbc}
\end{equation}
$\wh V_a^{\rm D5}$ is the restriction of $\wh V^{\rm D5}$ to the face $a$.
$V_\mu^{\rm NS5}$ is similarly defined as the restriction of
$V^{NS5}$ to the face $\mu$.
In order to satisfy this condition, we need non-vanishing
$V^{\rm NS5}$.
Let us introduce the following gauge field on the NS5-brane
depending on the $\U(1)_{\rm B}$ gauge field $V$:
\begin{equation}
V^{\rm NS5}=\sum_\mu {\cal B}V,
\label{ans}
\end{equation}
where ${\cal B}$ is the function on the NS5-brane worldvolume
defined by
\begin{equation}
{\cal B}=b_\mu\delta(\mu).
\end{equation}
The boundary condition (\ref{abmnbc}) requires
the coefficients $b_\mu$ satisfy (\ref{qbb}).
Namely, combining two equations
(\ref{qbb}) and (\ref{qbys}),
we can show that the boundary condition
(\ref{abmnbc}) is satisfied.

\begin{figure}[htb]
\epsfxsize=7cm
\centerline{\epsfbox{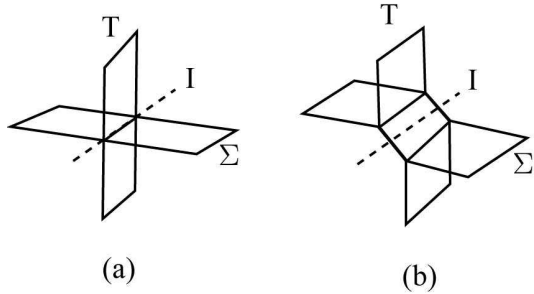}}
\caption{(a) $N$ D5-branes $T$ and NS5-brane $\Sigma$ intersecting on edge $I$.
(b) $N$ D5-branes $T$ and NS5-brane $\Sigma$ sharing strips around edge $I$.}
\label{defx.eps}
\end{figure}

This boundary condition can also be explained by the
cancellation of divergence of field strength on the D5-branes
and NS5-brane.
The field strengths for the potentials (\ref{vd5baryon}) and
(\ref{ans}) are
\begin{equation}
\wh{\cal F}^{\rm D5}=
{\cal S}dV{\bf 1}_N
+{\cal Q}\wedge V{\bf 1}_N.
\end{equation}
\begin{equation}
{\cal F}^{\rm NS5}={\cal B}dV+d{\cal B}\wedge V
\end{equation}
(Now we assume the vanishing $B_2$ and $C_2$.)
These field strengths have the terms
which include $V$ without derivative,
and they induce a mass for the four-dimensional gauge field $V$.
Fortunately, even though $\wh{\cal F}^{\rm D5}$ and ${\cal F}^{\rm NS5}$
live on different branes,
the unwanted terms in these field strengths
cancel each other when the relations
(\ref{qbb}) and (\ref{qbys})
hold.
To show this, let us
deform the 4-junction along the edge $I$
to two 3-junctions connected by $(N,1)$ fivebrane.
(Figure \ref{defx.eps})
We can regard this $(N,1)$ fivebrane as a superposition
of NS5-brane and $N$ D5-branes,
and the field strength ${\cal F}^{(N,1)}$ on
the $(N,1)$ fivebrane is the sum of
$\tr\wh{\cal F}^{\rm D5}$ and ${\cal F}^{\rm NS5}$.
Because the unwanted term in the field strengths
have support in the $(N,1)$ fivebrane,
they are canceled
if the relations (\ref{qbb}) and (\ref{qbys}) hold.

The gauge field (\ref{ans}) on the NS5-brane
couples to D-strings ending on the NS5-brane.
As we mentioned above, the baryonic operator
$\det\Phi_I$ corresponds to the D-string stretched between
faces $\mu$ and $\nu$ on the NS5-brane sharing the edge $I$.
The relation (\ref{qbb}) guarantees that the charge of
the baryonic operator is $N$ times the charge of $\Phi_I$.

The gauge field on NS5-brane also plays another important role.
If it were absent,
the gauge field $V$ would live
only in the compact manifold (D5-brane world volume),
and it would become dynamical field with the coupling constant of the same order
as the gauge coupling constant of $\SU(N)$ gauge groups.
Due to the gauge fields on NS5-brane,
the non-compactness of the NS5-brane
worldvolume provides the infinitely large volume factor in
the kinetic term of the gauge field $V$, and
it becomes non-dynamical gauge field
corresponding to a global symmetry.

Now we can interpret the prescription
for obtaining baryonic charge assignments.
The parameters introduced in (\ref{bba}) are
regarded as the parameters in the relation (\ref{ans}),
and (\ref{qbb}) is physically interpreted as
the boundary condition imposed on the gauge field
on the NS5-brane.
We can also derive the constraint (\ref{vcondition})
by combining two equations
(\ref{qbb}) and (\ref{qbys}).
Because the zig-zag path
$\bm\mu$ is the boundary of the
face $\mu$ on the NS5-brane,
the relation (\ref{qbb}) implies that
the one-form ${\cal Q}$ is given by
\begin{equation}
{\cal Q}=\sum_\mu b_\mu\delta({\bm\mu}).
\end{equation}
As (\ref{qbys}) shows, ${\cal Q}$ is exact on $T$ and
belongs to the trivial cohomology class.
This is equivalent to the relation (\ref{vcondition}).

\subsection{$\theta$ angles and $\U(1)_{\rm B}\SU(N_a)^2$ anomalies}
The charge assignments we discussed above
give anomaly free $\U(1)_{\rm B}$ symmetries.
The existence of the $\U(1)\SU(N_a)^2$ anomaly for a $\U(1)$ symmetry
implies that
the $\SU(N_a)$ $\theta$-angle is shifted by the
$\U(1)$ rotations.
In this subsection, we identify parameters in the fivebrane system
which correspond to the $\theta$ angles,
and the relations (\ref{qbb}) and (\ref{qbys})
guarantee the invariance of the parameters under the $\U(1)$ rotations.

Let $\delta_4(a)$ be the four-form $\delta$ function
in ten-dimensional spacetime
supported on the worldvolume of the D5-branes on the face $a$,
whose boundary is the cycle ${\bm a}$ on the NS5-brane $\Sigma$.
If the $\U(N)$ gauge field $\wh V_a^{\rm D5}$ on the D5-branes
does not vanish,
D3-current $(1/2\pi)\delta_4(a)\wedge\tr\wh F_a^{\rm D5}$ and D1-current
$(1/8\pi^2)\delta_4(a)\wedge\tr(\wh F_a^{\rm D5}\wedge\wh F_a^{\rm D5})$
are carried by the D5-branes
in addition to the D5-current $N\delta_4(a)$.
The D1-current on the D5-branes on the face $a$
electrically couples to the $\U(1)$ gauge field $V^{\rm NS5}$
on the NS5-brane $\Sigma$:
\begin{equation}
S=\frac{1}{8\pi^2}\int_{{\bf R}^4\times{\bm a}} V^{\rm NS5}
\wedge\tr(\wh F_a^{\rm D5}\wedge\wh F_a^{\rm D5})
\end{equation}
This implies that the $\theta$-angle for the $\SU(N_a)$ gauge group
is given by
\begin{equation}
\theta_a=\oint_{\bm a}V^{\rm NS5}.
\label{thetaangle}
\end{equation}
(There is also the RR $2$-form potential contribution to the $\theta$-angles,
which is omitted here.)
On the other hand, the magnetic coupling of $V^{\rm NS5}$ to
the D3-current is represented by the Bianchi
identity
\begin{equation}
dF^{\rm NS5}+\sum_b\delta({\bm b})\wedge\tr\wh F^{\rm D5}_b=0.
\end{equation}
This implies that the field strength $F^{\rm NS5}$ is given by
\begin{equation}
F^{\rm NS5}=dV^{\rm NS5}-\sum_b\delta({\bm b})\tr\wh V_a^{\rm D5},
\end{equation}
and the gauge invariance of this field strength
requires the gauge field $V^{\rm NS5}$ be transformed
by the gauge transformation $\delta V^{\rm D5}=d\wh\lambda^{\rm D5}$
as
\begin{equation}
\delta V^{\rm NS5}=\sum_b\delta({\bm b})\tr\wh \lambda_b^{\rm D5}.
\label{gaugeonb}
\end{equation}
If the baryonic symmetry is realized as gauge symmetry on D5-branes
by the embedding (\ref{vd5baryon}),
the transformation parameters $\wh\lambda_b$
in (\ref{gaugeonb})
is related to the parameter $\lambda$ for the baryonic symmetry
by
\begin{equation}
\wh\lambda_a^{\rm D5}={\bf 1}_NS_a\lambda,
\end{equation}
and this gauge transformation
changes the $\theta$-angle (\ref{thetaangle}) by
\begin{equation}
\delta\theta_a
=\sum_b\oint_{\bm a}\delta({\bm b})\tr\wh\lambda_a^{({\rm D5})}
=N\left\langle{\bm a},\sum_bS_b{\bm b}\right\rangle\lambda.
\label{u1anom}
\end{equation}

We can show that this anomaly cancels if the relations
(\ref{qbb}) and (\ref{qbys}) are satisfied.
From the relations (\ref{qbb}) and (\ref{qbys})
we can show that the $1$-chain
$\sum_bS_b{\bm b}$
on $\Sigma$ can be represented as the boundaries of the faces on
the $\ol\Sigma$.
Therefore, this is homologically trivial:
\begin{equation}
\sum_a S_a{\bm a}=\sum_\mu b_\mu\partial\mu=0.
\quad\mbox{on $\ol\Sigma$}.
\label{scond}
\end{equation}
The relation (\ref{scond})
guarantees the cancellation of
the anomaly (\ref{u1anom}).

\subsection{Mesonic symmetries}
As we mentioned in \S\S\ref{strandop.ssec},
we can associate gauge invariant mesonic operators
to closed strings.
Thus mesonic symmetries coupling to some of them
should be realized with the NS-NS $2$-form field, $B$.
For a charge assignment $Q_I$, we can give the 
gauge field
$V$ coupling to the bi-fundamental field with charge $Q_I$ as
\begin{equation}
B={\cal Q}\wedge V+fdV
\label{imprb}
\end{equation}
where $f$ is a function on $T$ such that
${\cal Q}-df$ is a smooth function.
The second term is introduced to avoid the
divergence of the energy induced by the field strength $H_3=dB$.
We need to solve equations of motion to determine the function $f$.
We, however, do not need the precise form of $f$.

With the existence of the D5-branes
wrapped on $T$,
the above $B$ field appear in the D5-brane action
through the field strength
$\wh{\cal F}^{\rm D5}=d\wh V^{\rm D5}+B{\bf 1}_N$,
and the first term in 
(\ref{imprb}) induce the mass term
for the gauge field $V$ through the field strength.
This problem can be avoided by introducing the
gauge field on the NS5-brane just in the same way as the
baryonic case.
We again introduce the gauge field (\ref{ans}) on the NS5-brane,
and require the cancellation between unwanted terms
in $\wh{\cal F}^{D5}$ and ${\cal F}^{NS5}$.
This cancellation is realized if the relation
(\ref{qbb}) is satisfied.
The difference from the baryonic case is that
the closed one-form ${\cal Q}$ does not have to be exact and
the condition (\ref{qbys}) is not imposed.
This fact corresponds to the fact that the condition
(\ref{vcondition}) is not imposed on the parameters $b_\mu$
when we determine mesonic charge assignments
by following the prescription
we mentioned above.

If ${\cal Q}$ is exact and given by (\ref{qds}),
it gives a baryonic charge assignments.
We can use the above realization with the $B$ field
in this case, too.
Thus, we have two ways to realize the baryonic symmetries.
These two are actually gauge equivalent through $B$-field
gauge transformation.
For an exact ${\cal Q}$,
we can adopt the function $f={\cal S}$ in (\ref{imprb}),
and the $B$-field is given by
\begin{equation}
B={\cal Q}\wedge V+{\cal S}dV=d({\cal S}V)
\label{exactb}
\end{equation}
This choice of the function $f$ minimizes
the energy of the NS-NS field $H$ in the bulk.
If we perform the $B$-field gauge transformation
\begin{equation}
\delta B=d\Lambda,\quad
\delta\wh V^{\rm D5}=-\Lambda{\bf 1}_N
\end{equation}
with the parameter $\Lambda=-{\cal S}V$,
the $B$ field becomes zero.
Instead, the gauge field on D5-branes
becomes non-vanishing and given by (\ref{vd5baryon}).

We always have the ambiguity
associated with this equivalence when we realize a given symmetry
as gauge symmetry in the brane configuration.
This can be identified with the
mixing ambiguity between the mesonic symmetries
and baryonic symmetries.

\subsection{R-symmetry}\label{rsym.sec}
The prescription for obtaining
R-charge assignment $R_I$ is similar to the
prescription for flavor symmetries\cite{Butti:2005vn}.
Instead of the parameters satisfying (\ref{condona1}),
we associate with vertices in the toric diagram
the parameters $a_\mu$ satisfying
\begin{equation}
\sum_{\mu=1}^d a_\mu=2.
\label{suma2}
\end{equation}
For a set of parameters $a_\mu$ 
satisfying (\ref{suma2}), the R-charge of $\Phi_I$ is given by
\begin{equation}
R_I=\sum_{\rho=\nu}^{\mu-1}a_\rho\quad
\mbox{for $\sign(\mu,\nu)>0$},\quad
R_I=\sum_{\rho=\mu}^{\nu-1}a_\rho\quad
\mbox{for $\sign(\mu,\nu)<0$},
\label{rdef}
\end{equation}
where $\bm\mu$ and $\bm\nu$ are the two zig-zag paths
sharing the edge $I$.

As in the case of baryonic and mesonic symmetries,
it is convenient to associated with the edges of the toric diagram
the parameters $\phi_\mu$
which give $a_\mu$ as the differences among them.
However,
the relation (\ref{suma2}) implies that
$\phi_\mu$ cannot be
single-valued parameters.
Thus we define them as angular parameters defined modulo $2\pi$.
The relation between $a_\mu$ and $\phi_\mu$ is
\begin{equation}
\phi_\mu-\phi_{\mu-1}=\pi a_\mu\quad\mod 2\pi.
\end{equation}
The equations in (\ref{rdef})
for R-charges are rewritten as
\begin{equation}
\pi R_I=\sign(\mu,\nu)(\phi_\mu-\phi_\nu)\quad
\mod 2\pi,
\end{equation}
where $\bm\mu$ and $\bm\nu$ are the two zig-zag paths
sharing the edge $I$.

In order to fix the $2\pi$ ambiguity, we need to
carefully define the difference of two angles.
For this purpose, we assume that all the R-charges of bi-fundamental fields
satisfy
\begin{equation}
0\leq R_I\leq 1.
\label{unitarity}
\end{equation}
We restrict our attention to R-symmetry
satisfying this condition.

If we assume that the condition $-1\leq R_I\leq 1$, which is looser than
(\ref{unitarity}), is satisfied,
we can fix the mod $2\pi$ ambiguity by
\begin{equation}
R_I=\frac{1}{\pi}\sign(\mu,\nu)[[\phi_\mu-\phi_\nu]],
\label{rphi}
\end{equation}
where $[[x]]$ is defined by
\begin{equation}
[[x]]\equiv x\mod 2\pi,\quad
-\pi\leq [[x]]\leq \pi.
\end{equation}
The positivity of the R-charge in (\ref{rphi})
requires $\phi_\mu$ satisfy
\begin{equation}
\sign(\mu,\nu)[[\phi_\mu-\phi_{\mu-1}]]\geq0.
\end{equation}
With this inequality, we can show that
the cyclic order of $\phi_\mu$
is the same with the order of the external legs of web diagram
(if we define the order of degenerate angles appropriately).

R-charges satisfying (\ref{unitarity}) can be
described by an isoradial embedding of
the bipartite graph\cite{Hanany:2005ss}.
In such embeddings, each edge in the tiling
is represented as a diagonal of rhombus.
A zig-zag path $\mu$ is represented as rhombi path
and we can identify the angle $\phi_\mu$ to
the direction of the sides of rhombi inside the rhombi path.
In such rhombus lattice,
the R-charges are represented as internal angles of the
rhombi\cite{Hanany:2005ss}, and are given by (\ref{rphi}).

Because angles $\phi_\mu$ are associated with the
external legs of a web diagram,
and the cyclic order of $\phi_\mu$ is the same with that of
the external legs under the assumption (\ref{unitarity}),
it is natural to interpret the parameters $\phi_\mu$
as the direction of external lines of web diagrams on the
$46$-plane.
In order to confirm this speculation,
let us compute R-charges of charged objects
in the brane configuration.

Let $V_R$ be the gauge field for the R-symmetry.
We compute the coupling of $V_R$ to D-strings
stretched between two faces on $\Sigma$, which correspond to the baryons.

We should first identify the gauge field $V_R$.
Because the R-symmetry is the $89$-rotation of the system
in the brane realization, the gauge field $V_R$ enters in the metric
as
\begin{equation}
ds^2
=
ds_6^2
+dr^2
+r^2d\theta^2
+r^2\cos^2\theta d\phi^2
+r^2\cos^2\theta (d\psi+2V_R)^2
\end{equation}
when $V_R$ is pure gauge.
$ds_6^2$ is the flat metric for $012357$ directions.
For the codimensions $4689$ of the D5-branes, we introduce
the polar coordinates by
\begin{equation}
x^4+ix^6=re^{i\phi}\cos\theta,\quad
x^8+ix^9=re^{i\psi}\sin\theta.
\end{equation}
We here normalize the gauge field $V_R$ so that
$x^8+ix^9$ has charge $2$.
This means that the R charge of supercharge,
which has spin $J_{89}=1/2$ on the 89 plane,
is $1$.
If $F_R\equiv dV_R\neq0$,
the metric should be modified by the
corresponding curvature.
However, we do not need the $F_R$ correction
because the R-charge is determined only
by the minimal coupling
of $V_R$ to charged objects.

Let us consider the RR $3$-form flux induced by the $N$ D5-branes wrapped
on $T$, the torus along $x^5$ and $x^7$.
(See Table \ref{brane.tbl}.)
If $V_R=0$ the flux is given by
\begin{equation}
G_3=\frac{N}{\pi}\sin\theta\cos\theta d\theta\wedge d\phi\wedge d\psi.
\end{equation}
The gauging of the $89$-rotation
can be taken into account
by the replacement
\begin{equation}
d\psi\rightarrow d\psi+2V_R.
\label{rtwist}
\end{equation}
By this replacement
we obtain the $V_R$ dependence of the flux $G_3$
as
\begin{equation}
\mbox{$V_R$ dependent part of $G_3$}
=\frac{2N}{\pi}\sin\theta\cos\theta d\theta\wedge d\phi\wedge V_R.
\end{equation}
The corresponding $2$-form potential
is
\begin{equation}
\mbox{$V_R$ dependent part of $C_2$}
=\frac{N}{\pi}\cos^2\theta d\phi\wedge V_R.
\end{equation}
(We assumed the gauge field $V_R$ varies slowly and
we neglect the $dV_R$ term.)
We chose the gauge such that the $\U(1)_R$ isometry is
manifest and the potential is non-singular except at the
origin of the $4689$ space.

The coupling of $V_R$ to a baryon
are obtained by integrating this potential
along the worldvolume of the D-string corresponding to
the baryon.
If $I$ is an edge shared by two zig-zag paths $\bm\mu$ and $\bm\nu$,
the D-string corresponding to the baryon $\det\Phi_I$ is
stretched between external legs $\mu$ and $\nu$.
If we assume that (the time slice of) the worldvolume of the D-string
is a curve on the $46$-plane
connecting the legs $\mu$ and $\nu$,
the coupling of this RR field to
D-string is given by
\begin{equation}
S
=\frac{N}{\pi}\sign(\mu,\nu)[[\phi_\mu-\phi_\nu]]\int V_R.
\label{d1vr}
\end{equation}
We assumed that the D1 worldvolume is a curve on the $46$-plane
with smallest $|\int d\phi|$,
and it does not make a detour around the origin.
The orientation of the D1-brane is chosen so that it gives
the correct orientation of open fundamental strings
through the process in Figure \ref{hw.eps}.
The coupling (\ref{d1vr}) shows that the R-charge
for $\Phi_I$ is given by (\ref{rphi}).

We should note that the argument above is not a proof of (\ref{rphi})
but a heuristic explanation which seems to support the relation
(\ref{rphi}).
Indeed, we cannot obtain different R-charges of component fields
in a supermultiplet in such a classical analysis.
For rigorous proof of the relation (\ref{rphi})
we should quantize open strings.

Another circumstantial evidence for the relation (\ref{rphi}) is
the fact that as we demonstrate below
we can obtain $\U(1)_{\rm R}\U(1)_{\rm B}^2$ 't~Hooft anomalies
which is consistent with the charge assignment (\ref{rphi})
by using the classical brane action.

\section{'t~Hooft anomalies}\label{anomaly.sec}
In this section, we discuss how some 't~Hooft anomalies
are reproduced by using the classical action of the brane system.

The global symmetries of gauge theories are realized as gauge symmetries
in string theory as we discussed above.
The anomalies associated with the symmetries
must locally cancel for the consistency of the theory.
This is achieved by so-called anomaly inflow mechanism\cite{Callan:1984sa,Naculich:1987ci,Blum:1993yd}.

An anomaly localized in a sub-manifold of the spacetime (branes, intersection of branes etc.)
causes
violation of the conservation law for the
current on the sub-manifold.
If we assume that the total theory is anomaly-free,
this violation must be compensated by an inflow of the current
from the ambient to the sub-manifold.
Using this fact,
we can compute anomalies in a sub-manifold as
inflows of the currents through the boundary of the system.

Let us see how this mechanism works in a simple example\cite{Green:1996dd}.
We consider an intersecting D-brane system
consisting of a D5-brane (D5$_A$) along 012345 and another D5-brane (D5$_B$)
along 016789 in type IIB theory.
We here consider the anomaly associated with the $\U(1)$ gauge symmetry on D5$_A$,
and assume that the gauge field on D5$_B$ vanishes for simplicity.
There is a Weyl fermion living on the intersection
and it couples to the gauge field on D5$_A$.
The anomaly arising in the intersection $I$ is
\begin{equation}
\delta\Gamma=\frac{1}{4\pi}\int_I\lambda F
\end{equation}
for the gauge transformation $\delta V=d\lambda$,
where $F=dV$ is the field strength on D5$_A$.
This anomaly at the intersection is locally canceled by the variation
of the Chern-Simons term of D5$_A$
\begin{equation}
S_{\rm CS}=\frac{1}{8\pi^2}\int_{{\rm D5}_A} G_3\wedge V\wedge F.
\end{equation}
The variation of this action is
\begin{equation}
\delta S_{\rm CS}=\frac{1}{8\pi^2}\int_{{\rm D5}_A} dG_3\wedge F\lambda
+\frac{1}{8\pi^2}\oint_{\partial{\rm D5}_A} G_3\wedge F\lambda,
\end{equation}
where the second term is the boundary term arises when we take the integral by part.
Because $G_3$ magnetically couples to D5$_B$ and
$dG_3=2\pi\delta_4({\rm D5}_B)$,
the first term cancels the anomaly at the intersection.
Instead, the anomaly appears in the second term as the boundary
term.
By using this, it is possible to compute the anomaly as the boundary term without using
knowledge of the intersection
on which the anomaly arises.

In the rest of this section,
we discuss three classes of 't~Hooft anomalies.
It is not clear if all the 't~Hooft anomalies
can be obtained as the boundary term of the variations of
the brane action
because there is possibility that anomalies are
canceled by variations of the bulk action of supergravity.
In the following we simply assume that
the anomalies are
locally canceled by boundary terms of variations
of the brane action,
and we show that the correct anomalies are obtained for
$\tr(\U(1)_{\rm M}\U(1)_{\rm B}^2)$ and
$\tr(\U(1)_{\rm R}\U(1)_{\rm B}^2)$.
(It will turn out that we need to take account of bulk action
to obtain a gauge independent result.)

\subsection{$\U(1)_{\rm B}^3$}
We first consider the $\tr\U(1)_{\rm B}^3$ anomalies.
It is known that these anomalies always vanish.
\cite{Butti:2005vn}

Let $b_\mu^i$ ($i=1,2,3$) be
three sets of parameters
which give
three $\U(1)_{\rm B}$ charge assignments $B_I^i$ by the relation (\ref{qbb}).
Namely,
the parameters $b_\mu^i$ satisfy
\begin{equation}
\sum_\mu b_\mu^i{\bm\mu}=0,
\label{bmuiconst}
\end{equation}
and the charges $B_I^i$ are given by
\begin{equation}
B_I^i=\frac{1}{N}\sign(\mu,\nu)(b_\mu^i-b_\nu^i).
\label{baryonicbb}
\end{equation}
The coefficient of the $\tr\U(1)_{\rm B}^3$ anomaly is given by
\begin{equation}
\tr(B^1B^2B^3)\equiv
N^2\sum_IB_I^1B_I^2B_I^3.
\end{equation}
Using the expression
(\ref{baryonicbb}) for the charges,
this can be rewritten as
\begin{equation}
\tr(B^1B^2B^3)
=\frac{1}{2N}\sum_{\mu,\nu}\langle{\bm\mu},{\bm\nu}\rangle
(b^1_\mu-b^1_\nu)
(b^2_\mu-b^2_\nu)
(b^3_\mu-b^3_\nu).
\label{qqqbbb}
\end{equation}
We used the fact that we can replace the
summation with respect to the indices $I$ by the
summation over pairs of cycles $(\mu,\nu)$
with multiplicities $|\langle{\bm\mu},{\bm\nu}\rangle|$:
\begin{equation}
\sum_I(\cdots)
=\frac{1}{2N}\sum_{\mu,\nu}|\langle{\bm\mu},{\bm\nu}\rangle|(\cdots).
\label{multi}
\end{equation}
When the summand $(\cdots)$ is $1$,
this relation gives the formula (\ref{nmatter}).
After expanding the right hand side in
(\ref{qqqbbb}),
we have eight terms cubic with respect to $b_\mu^i$.
It is easy to show that all these eight terms vanish separately.
For example,
$(1/2N)\sum_{\mu,\nu}\langle{\bm\mu},{\bm\nu}\rangle b_\mu^1b_\mu^2b_\mu^3$
vanishes due to (\ref{totalmu}), and
$(1/2N)\sum_{\mu,\nu}\langle{\bm\mu},{\bm\nu}\rangle b_\mu^1b_\mu^2b_\nu^3$
also vanishes due to (\ref{bmuiconst}).
As the result,
we obtain
\begin{equation}
\tr(B^1B^2B^3)=0.
\end{equation}
It is also easy to show $\tr B^i=0$.

We use this fact for the purpose of
fixing the ambiguity of
regularizations in the gauge theory and
total derivative terms in the Lagrangian of branes.
Namely, on the gauge theory side,
we choose regularization which does not break the $\U(1)_B$ symmetries,
and on the string theory side we use the brane action which does not produce boundary terms
when gauge transformations on branes corresponding to
$\U(1)_B$ symmetries are carried out.
This is the case if the action includes the $\U(1)$ gauge fields
on branes only through the gauge invariant field strengths.
For example, we use the following Chern-Simons term of D-branes,
which is manifestly gauge invariant:
\begin{equation}
S_{\rm CS}=\int C\wedge e^{(F-B)/2\pi}.
\label{scs}
\end{equation}

\subsection{$\U(1)_{\rm M}\U(1)_{\rm B}^2$}
Let $M_I$ and $B_I^i$ be a mesonic charge assignment and
baryonic charge assignments, respectively.
The baryonic charges $B_I^i$ are given by (\ref{baryonicbb})
with the parameters $b_\mu^i$ constrained by
(\ref{bmuiconst}).
The mesonic charges $M_I$ are given by
\begin{equation}
M_I=\frac{1}{N}\sign(\mu,\nu)(m_\mu-m_\nu),
\label{fff}
\end{equation}
where $m_\mu$ are parameters without constraint.

Using the expression
(\ref{baryonicbb}) and (\ref{fff})
for the charges
and the relation (\ref{multi}) for the multiplicity,
we obtain the following anomaly coefficient for $\tr(\U(1)_{\rm M}\U(1)_{\rm B}^2)$:
\begin{eqnarray}
A_{Mij}
&\equiv&\tr(MB^iB^j)
\nonumber\\
&\equiv&N^2\sum_IM_IB_I^iB_I^j
\nonumber\\
&=&\frac{1}{2}\sum_{\mu,\nu}
\langle{\bm\mu},{\bm\nu}\rangle
(m_\mu-m_\nu)
(b_\mu^i-b_\nu^i)
(b_\mu^j-b_\nu^j)
\nonumber\\
&=&\sum_{\mu,\nu}
m_\mu\langle{\bm\mu},{\bm\nu}\rangle
b_\nu^ib_\nu^j
\end{eqnarray}

The existence of the non-vanishing coefficient
$A_{Mij}$ means that
under
a gauge transformation
\begin{equation}
\delta V_M=d\lambda_M,\quad
\delta V_B^i=d\lambda_B^i,
\label{vfvbgauge}
\end{equation}
the effective action is not invariant.
Let $\delta\Gamma$ be the variation of the effective action
under the gauge transformation (\ref{vfvbgauge}).
It is important that $\delta\Gamma$ depends on the regularization of loop amplitudes.
If we use a regularization
in which the three vertices in
the triangle fermion loop
are treated symmetrically,
we obtain the following variation:
\begin{equation}
\delta\Gamma
=\frac{1}{24\pi^2}A_{Mij}\lambda_M dV_B^i\wedge dV_B^j
+\frac{1}{12\pi^2}A_{Mij}\lambda_B^i dV_M\wedge dV_B^j
\end{equation}
We may use other regularizations different
by finite counter terms.
For example, we can add the following counter term.
\begin{equation}
S_{\rm counter}=-\frac{1}{12\pi^2}A_{Mij}V_M\wedge V_B^i\wedge dV_B^j
\end{equation}
In this case, we have the following anomaly
for the gauge transformation (\ref{vfvbgauge})
\begin{equation}
\delta(\Gamma+S_{\rm counter})=\frac{1}{8\pi^2}A_{Mij}\lambda_M dV_B^i\wedge dV_B^j.
\label{anomgauge}
\end{equation}
This variation does not include the $\U(1)_B$ gauge transformation parameters
$\lambda_B^i$.
As we mentioned in the previous subsection,
we here use the $\U(1)_B$ invariant regularization.
Thus we should adopt
(\ref{anomgauge}) as the anomaly which we compare to the variation of the
brane action.
This regularization corresponds to
the action in which gauge fields on branes appear only through the
gauge invariant field strengths.
We can of course use a different regularization,
and it corresponds to a different choice of boundary terms
in the brane action.

The NS5-brane action has the term
\begin{equation}
S_{\rm NS5}=\frac{1}{8\pi^2}\int_{{\bf R}^4\times\Sigma}B\wedge dV^{\rm NS5}\wedge dV^{\rm NS5}.
\label{bffterm}
\end{equation}
For the mesonic charge assignment (\ref{fff}),
the corresponding gauge field $V_M$
enters in the $B$-field as
\begin{equation}
B
=\sum M_I\delta(I)\wedge V_M+fdV_M
=\sum m_\nu\delta({\bm\nu})\wedge V_M+fdV_M.
\label{bff}
\end{equation}
The second term in (\ref{bff}) is gauge invariant and
only the first term contribute the boundary term in the variation
of the action.
Under the $\U(1)_M$ gauge transformation
$V_M=d\lambda_M$,
the $B$-field (\ref{bff})
is transformed by
\begin{equation}
\delta B=\sum_\nu m_\nu \delta({\bm\nu})\wedge d\lambda_M.
\label{btrgauge}
\end{equation}
The action
(\ref{bffterm}) is gauge
invariant only up to the boundary term,
and the 
transformation (\ref{btrgauge}) produces the
boundary term
\begin{equation}
\delta S_{\rm NS5}
=\frac{1}{8\pi^2}
\int_{{\bf R}^4\times\partial\Sigma}\lambda_M
\sum_\nu m_\nu \delta({\bm\nu})
\wedge dV^{\rm NS5}
\wedge dV^{\rm NS5}.
\end{equation}
The NS5-brane has $d$ boundaries labeled by $\mu$,
which are the $\bm\mu$ cycles in the torus.
Therefore,
\begin{eqnarray}
\delta S_{\rm NS5}
&=&\frac{1}{8\pi^2}
\int_{{\bf R}^4}\lambda_M\sum_\nu\int_\nu
\left(\sum_\mu m_\mu \delta({\bm\mu})\right)
\wedge\left(\sum_i b_\nu^i dV_B^i\right)
\wedge\left(\sum_j b_\nu^j dV_B^j\right)
\nonumber\\
&=&\frac{1}{8\pi^2}
\int_{{\bf R}^4}\lambda_M\sum_{i,j}\sum_{\mu,\nu}
m_\mu\langle{\bm\mu},{\bm\nu}\rangle b_\nu^i b_\nu^j
dV_B^i
\wedge dV_B^j.
\end{eqnarray}
where we used (\ref{deltaint}).
The final expression is the same with the anomaly (\ref{anomgauge}).

\subsection{$\U(1)_R\U(1)_B^2$}\label{rbb.ssec}
On the gauge theory side,
the anomaly coefficient for $\tr(\U(1)_R\U(1)_B^2)$ is
\begin{equation}
A_{Rij}=\tr(RB^iB^j)\equiv N^2\sum_I(R_I-1)B_I^iB_I^j.
\end{equation}
The trace is taken over all the fermions in the theory,
and the R-charge of the fermion in the chiral multiplet $\Phi_I$
is given by
\begin{equation}
R_I-1=\sign(\mu,\nu)\frac{1}{\pi}[[\phi_\mu-\phi_\nu-\pi]].
\end{equation}
Therefore, the anomaly coefficient is
\begin{equation}
A_{Rij}=\frac{1}{2\pi}\sum_{\mu,\nu}\langle{\bm\mu},{\bm\nu}\rangle
[[\phi_\mu-\phi_\nu-\pi]]
(b_\mu^i-b_\nu^i)
(b_\mu^j-b_\nu^j).
\end{equation}
The variation of the effective action
computed with the $\U(1)_B$ invariant regularization
is
\begin{eqnarray}
\delta\Gamma
&=&\frac{1}{8\pi^2}\sum_{i,j}A_{Rij}\int_{{\bf R}^4} \lambda_RF_B^i\wedge F_B^j
\nonumber\\
&=&\frac{1}{16\pi^2}\sum_{\mu,\nu}
\langle{\bm\mu},{\bm\nu}\rangle
\frac{1}{\pi}[[\phi_\mu-\phi_\nu-\pi]]
\nonumber\\&&\quad\times
\int_{{\bf R}^4} \lambda_R(F^{\rm NS5}_\mu-F^{\rm NS5}_\nu)\wedge(F^{\rm NS5}_\mu-F^{\rm NS5}_\nu).
\label{eq80}
\end{eqnarray}
In (\ref{eq80}) we rewrote the anomaly in terms of gauge fields on
the NS5-branes.

\subsubsection{The conifold theory}
To illustrate how the anomaly (\ref{eq80})
 is reproduced in the brane system,
we first analyze the conifold theory as the simplest example.
It is $\SU(N)^2$ gauge theory with four bi-fundamental chiral multiplets $A_1$, $A_2$, $B_1$ and $B_2$.
The tiling has two faces and the chiral multiplets
are coupled to the $\U(N)_a=\SU(N)_a\times\U(1)_a$ ($a=1,2$) gauge fields
on the D5-branes with charges given in Table \ref{coni.tbl}.
\begin{table}[htb]
\caption{The matter contents of the conifold theory.}
\label{coni.tbl}
\begin{center}
\begin{tabular}{cccccc}
\hline
\hline
          & $\SU(N)_1$ & $\SU(N)_2$ & $\U(1)_1$ & $\U(1)_2$ & $\U(1)_R$ \\
\hline
$A_{1,2}$ & $\bf N$  & $\ol{\bf N}$ & $1$ & $-1$ & $1/2$ \\
$B_{1,2}$ & $\ol{\bf N}$  & $\bf N$ & $-1$ & $1$ & $1/2$ \\
\hline
\end{tabular}
\end{center}
\end{table}
Only one combination of $\U(1)_1$ and $\U(1)_2$
symmetries couples to the matter fields,
and it is the $\U(1)_B$ symmetry of the conifold theory.
The $\U(1)_R\U(1)_B^2$ 't~Hooft anomaly is given by
\begin{equation}
\delta\Gamma=-\frac{N^2}{4\pi^2}\int\lambda_R F_B\wedge F_B,\quad
F_B=\frac{1}{N}\tr(\wh F_1^{\rm D5}-\wh F_2^{\rm D5}).
\end{equation}
where $\wh F_1^{\rm D5}$ and $\wh F_2^{\rm D5}$
are the $\U(N)$ gauge fields on
two faces, and $F_B$ is the field strength of the $\U(1)_B$ gauge field.

The gauge fields on the D5-branes are related to the gauge fields on the NS5-brane
by the boundary condition
\begin{equation}
F^{\rm NS5}_{45}-F^{\rm NS5}_{67}=\tr(\wh F_1^{\rm D5}-\wh F_2^{\rm D5})=NF_B,
\end{equation}
where $F^{\rm NS5}_{45}$ and $F^{\rm NS5}_{67}$ are the gauge fields
on NS5-branes along $45$ and $67$ directions, respectively.
By this relation, the anomaly computed in the gauge theory
can be rewritten in the following
$N$-independent form:
\begin{equation}
\delta\Gamma=-\frac{1}{4\pi^2}\int\lambda_R(F_{45}^{\rm NS5}-F_{67}^{\rm NS5})\wedge(F_{45}^{\rm NS5}-F_{67}^{\rm NS5}).
\label{conianom}
\end{equation}

Before showing that the classical action of
NS5-brane system actually reproduce this anomaly,
we give another interpretation of the anomaly
using another gauge theory.
Because (\ref{conianom}) is independent of $N$,
the system of the NS5-branes must have this anomaly even when the D5-branes
are absent.
Without D5-branes, the system preserves ${\cal N}=2$ supersymmetry,
and one hyper-multiplet arises from D-strings stretched between two NS5-branes.
We denote this hyper-multiplet as two chiral multiplets $Q$ and $\wt Q$.
These multiplets couple to the gauge fields on the NS5-branes
with charges given in Table \ref{qq.tbl}.
\begin{table}[htb]
\caption{The charges of chiral multiplets arising at the intersection of two NS5-branes.
The $\U(1)_R$ charges given in this table are charges for scalar components.
The charges for the fermion components are less than them by $1$.}
\label{qq.tbl}
\begin{center}
\begin{tabular}{cccc}
\hline
\hline
          & $\U(1)_{45}$ & $\U(1)_{67}$ & $\U(1)_R$ \\
\hline
$Q$     & $1$ & $-1$ & $0$ \\
$\wt Q$ & $-1$ & $1$ & $0$ \\
\hline
\end{tabular}
\end{center}
\end{table}
The 't~Hooft anomaly for this gauge theory is precisely
the same with the anomaly
(\ref{conianom}) for the conifold theory.

Let us reproduce this anomaly as the gauge transformation
of the classical NS5-brane action.
It is convenient to perform the S-duality transformation
to make the NS5-brane system to intersecting D5-branes.
Although this is simply the field redefinition, and does not change the physics at all,
it makes equations below somewhat simpler.

We can treat RR-fields in a unified way by using the formal sum
$G\equiv G_1+G_3+G_5+G_7+G_9$.
The electric-magnetic duality relation is
\begin{equation}
G={\ol *}G\equiv
-*G_9+*G_7-*G_5+*G_3-*G_1,\label{emrel}
\end{equation}
where ${\ol *}$ is the Hodge dual operator with
the alternating signature depending on the rank of fields.
We assume $G_1$ and $G_9$ vanish.
The relation between the field strength and the RR potential is given by
\begin{equation}
G=e^{B_2/2\pi}\wedge d(e^{-B_2/2\pi}\wedge C),
\label{strength}
\end{equation}
where $C=C_2+C_4+C_6$ is the formal sum of the RR potential fields.
The equations of motion and Bianchi identities for the RR fields with the presence of D5-branes are
packed in one equation
\begin{equation}
dG=\frac{1}{2\pi}H_3\wedge G+2\pi e^{B_2/2\pi}\wedge J,
\label{eomg}
\end{equation}
where $J$ is the D-brane current carried by D5-branes
\begin{equation}
J=\delta_4\wedge e^{-F^{\rm D5}/2\pi}.
\label{d5current}
\end{equation}
$\delta_4$ is the $4$-form delta function supported by the D5-brane worldvolumes.

For simplicity, we here neglect the NS-NS $2$-form field and
the self-duality condition for the RR field strengths
because they do not play
essential roles.
We take them into account when we discuss general case below.

In the brane system realizing the conifold theory
we have two D5-branes; one is along the 45 direction,
and the other is along the 67 direction.
Let us consider RR gauge fields induced by the D5-brane
along the 45 direction.
We first assume the $\U(1)_R$ gauge field
vanishes.
In this case, the four-form
$\delta_4$ representing the D5-brane worldvolume is
\begin{equation}
\delta_4({\rm D5}_{45})=
\delta(x^6)dx^6\wedge
dx^7\wedge
\delta(x^8)dx^8\wedge
\delta(x^9)dx^9.
\label{charge0}
\end{equation}
We assume that the compactification radius of $x^7$ is sufficiently small
and use the smeared charge density.
The equation (\ref{eomg}) becomes
\begin{equation}
dG=2\pi\delta_4\wedge e^{-F_{45}^{\rm D5}/2\pi}.
\label{geom}
\end{equation}
With the assumption of slow variation of $F_{45}^{\rm D5}$,
we can easily solve (\ref{geom}) with
\begin{equation}
G=\frac{1}{2}dx^7\wedge \sin\theta d\theta\wedge d\psi
\wedge e^{-F_{45}^{\rm D5}/2\pi},
\label{gmagnetic}
\end{equation}
where we define the following polar coordinates in the $689$ space:
\begin{equation}
x^6=r\cos\theta,\quad
x^8+ix^9=r\sin\theta e^{i\psi}.
\end{equation}
The corresponding RR potential is
\begin{equation}
C=\frac{1}{2}(\cos\theta+c)dx^7\wedge d\psi
\wedge e^{-F_{45}^{\rm D5}/2\pi},
\end{equation}
where $c$ is the integration constant which can be set arbitrarily.

Because the $\U(1)_R$ symmetry is the rotation along the coordinate $\psi$,
the introduction of the non-vanishing $\U(1)_R$ gauge field
can be achieved by replacing $d\psi$ by $d\psi+2V_R$.
As a result, we have the following RR potential:
\begin{equation}
C=\frac{1}{2}(\cos\theta+c)dx^7\wedge(d\psi+2V_R)\wedge e^{-F_{45}^{\rm D5}/2\pi}.
\label{RR45}
\end{equation}
Let us consider the coupling of this RR-field induced by D5$_{45}$
with the other D5-brane along $67$ direction.
The action describing the coupling is
\begin{equation}
S_{\rm D5}=\int_{{\rm D5}_{67}} C\wedge e^{F_{67}^{\rm D5}/2\pi},
\end{equation}
where $C$ is given by (\ref{RR45}).
By the $\U(1)_R$ gauge transformation $V_R=d\lambda_R$,
this action
produces the boundary term
\begin{equation}
\delta S_{\rm D5}
=\int_{\partial D5}\lambda_R(c+\cos\theta) dx^7\wedge e^{(F_{67}^{\rm D5}-F_{45}^{\rm D5})/2\pi}
\end{equation}
The D5-brane is a cylinder and has two boundaries.
Because these two have the opposite orientations,
the contribution of the integration constant $c$ cancels,
and we obtain
\begin{equation}
\delta S_{\rm D5}
=\frac{1}{4\pi^2}\int_{{\bf R}^4}\lambda_R
(F_{67}^{\rm D5}-F_{45}^{\rm D5})
\wedge(F_{67}^{\rm D5}-F_{45}^{\rm D5})
\end{equation}
This is precisely the same with the anomaly (\ref{conianom}), which we
obtained from the gauge theory.
(Because of the S-duality transformation we performed,
$F^{\rm NS5}_{45}$ and $F^{\rm NS5}_{67}$ are
replaced by
$F^{\rm D5}_{45}$ and $F^{\rm D5}_{67}$, respectively.)

One may think that one should consider
the gauge field induced by the D5$_{67}$ and
its coupling to D5$_{45}$.
However, it gives the same anomaly and actually
these two are one thing obtained by
two ways.
Taking both of them is double counting.
Thus we should not take account of both of them.

\subsubsection{General case}
Let us consider general case.
In order to simplify the problem,
we take the weak coupling limit
in which we can neglect the back reaction of energy density of
branes, and assume the background spacetime is flat
and the dilaton is constant near the boundary we compute the anomaly flow.
This assumption also allow us to treat the NS-NS gauge field $H_3$
as the background, because
$H_3$ decouples in the weak coupling limit $g_{\rm str}\rightarrow 0$ from the other fields in the
equation of motion
$d*H_3=(g_{\rm str}/2\pi)G_3\wedge G_5$.
Because there are $N$ NS5-branes wrapped on $T$,
the NS-NS gauge field strength and the corresponding potential
are
\begin{equation}
H_3=\frac{N}{\pi}\sin\theta\cos\theta d\theta\wedge d\phi\wedge d\psi,\quad
B_2=\frac{N}{2\pi}\cos^2\theta d\phi\wedge d\psi,
\label{h3b2}
\end{equation}
where we define the following polar coordinates in the 4689 space.
\begin{equation}
x_4+ix_6=r\cos\theta e^{i\phi},\quad
x_8+ix_9=r\sin\theta e^{i\psi}.
\label{h3b2polar}
\end{equation}
The potential $B_2$ is singular at $\theta=0$,
and there exist Dirac string-like singularity
\begin{equation}
H_3-dB_2=N\int_0^{2\pi} \delta_3(L_\phi)d\phi,
\end{equation}
where $L_\phi$ is the radial semi-infinite segment
in the $46$-plane
specified by an angle $\phi$ (Figure \ref{lphi2.eps}),
and $\delta_3(L_\phi)$ is the $3$-form delta function
in the 4689 space
supported by $L_\phi$.

In order to solve
the self-duality equation
(\ref{emrel}) and the equation of motion (\ref{eomg})
with the background (\ref{h3b2}),
we take the following ansatz for $G$:
\begin{equation}
G=(1+{\ol *})\left(G^{\rm mag}-2\pi e^{B_2/2\pi}\wedge X\right),
\label{anz0}
\end{equation}
where $X$ is a constant zero-form in $4689$ space and
$G^{\rm mag}$ is a two-form in $4689$ space.
These can be forms in $012357$ space, too.
We assume that $G^{\rm mag}$ takes the form
\begin{equation}
G^{\rm mag}=g\wedge d\psi,
\label{cond1}
\end{equation}
with $g$ being a one-form in $4689$ space,
and satisfies
\begin{equation}
d{\ol *}G^{\rm mag}=0.
\label{cond2}
\end{equation}
The field $G$ given by (\ref{anz0}) trivially satisfies the self-duality condition (\ref{emrel}).
Substituting the ansatz (\ref{anz0}) above into the equation of motion (\ref{eomg}),
we obtain
\begin{equation}
dG^{\rm mag}=2\pi J
    -(H_3-dB_2)\wedge X.
\label{eom2}
\end{equation}

For a D5-brane configuration described by a web-diagram with $d$ legs
the D-brane current $J$ in (\ref{d5current}) is given by
\begin{equation}
J=\sum_\mu e^{-F_\mu^{\rm D5}/2\pi}\wedge\delta({\bm\mu})\wedge\delta_3(L_{\phi_\mu}).
\label{j101}
\end{equation}
According to the argument in \S\S\ref{rsym.sec},
we assumed that the directions of the legs
on the 46-plane agree with the angles $\phi_\mu$,
which determine R-charges of the bi-fundamental fields.
Because $L_\phi$ has endpoint at origin in the 4689 space,
we have
\begin{equation}
d\delta_3(L_\phi)
=\delta_{4689}\equiv
\delta(x_4)
\delta(x_6)
\delta(x_8)
\delta(x_9)
dx^4\wedge
dx^6\wedge
dx^8\wedge
dx^9.
\end{equation}
For the consistency, the exterior derivative of the right hand side
in (\ref{eom2}) must vanish because the left hand side is closed.
From this condition, we obtain
\begin{equation}
dJ=NX\wedge\delta_{4689}.
\end{equation}
and $X$ is determined as
\begin{equation}
X=\frac{1}{8\pi^2N}\sum_\mu \delta({\bm\mu})\wedge F_\mu^{\rm D5}\wedge F_\mu^{\rm D5}.
\label{Xis}
\end{equation}
(This can be defined only when $N>0$.
If $N=0$, $J$ is conserved by itself and we can set $X=0$.)

The equation of motion (\ref{eomg}) reduces
to the following equation for $G^{\rm mag}$:
\begin{equation}
dG^{\rm mag}
=\int_0^{2\pi}d\phi 
\rho(\phi)
\wedge\delta_3(L_\phi)
\label{dg0}
\end{equation}
where $\rho(\phi)$ is the formal sum of the forms in the
$012357$ space defined by
\begin{equation}
\rho(\phi)=2\pi\sum_\mu
\delta(\phi-\phi_\mu)
 e^{-F_\mu^{\rm D5}/2\pi}\wedge\delta({\bm\mu})
-X
\end{equation}
Now we need to find the solution $G^{\rm mag}$ to
the equations (\ref{cond2}) and (\ref{dg0}) in the form (\ref{cond1}).

Because we assume $F$ is slowly varying along 0123 and $\rho(\phi)$
is approximately constant in the $0123$ space,
solving (\ref{cond2}) and (\ref{dg0}) is essentially the problem
determining the electro-magnetic field in the four-dimensional space
coupling to a conserved current.
This problem is solved in Appendix \ref{maxwell.sec},
and the result is
\begin{equation}
G^{\rm mag}=\frac{1}{2\pi}df\wedge d\psi
\label{gmagsol}
\end{equation}
where $f$ is the function in $4689$ space determined by solving the differential equation
given in Appendix \ref{maxwell.sec}.
The function $f$ satisfies the boundary condition
\begin{equation}
f|_{\theta=0}=m,
\end{equation}
where the function $m$ is defined by
\begin{equation}
dm(x^4,x^6)\wedge \delta(x^8)dx^8\wedge \delta(x^9)dx^9
=\int_0^{2\pi}d\phi 
\rho(\phi)
\wedge\delta_3(L_\phi),
\end{equation}
and the solution to this differential equation is
\begin{equation}
m(x^4,x^6)=\sum_\nu [[\phi-\phi_\nu-\pi]]
\delta({\bm\nu})\wedge e^{-F_\nu^{\rm D5}}+c,
\label{eq113}
\end{equation}
where $c$ is a integration constant,
which is one-form in $T$ and a formal sum of $0$, $2$, and $4$-forms
in $4689$ space.
We here assume $c=0$, and we comment on the $c$ independence of
the result at the end of this section.

For the purpose of obtaining anomaly flow,
we need to compute the coupling of RR potential and the D5-branes.
As we saw in the simplest example of the conifold case,
only the part of the RR potential which is written as $Fd\psi$
with a zero-form $F$ in $4689$ space contributes to the anomaly.
If $C^{\rm mag}$ denotes this part,
we can easily show that $G^{\rm mag}$ given in (\ref{gmagsol}) and
$C^{\rm mag}$ are related by
\begin{equation}
G^{\rm mag}=dC^{\rm mag}
\end{equation}
and we can easily determine the potential $C^{\rm mag}$ on the $46$-plane as
\begin{equation}
C^{\rm mag}|_{\theta=0}
=\frac{1}{2\pi}m(x^4,x^6)d\psi
=\frac{1}{2\pi}\sum_\nu [[\phi-\phi_\nu-\pi]]
\delta({\bm\nu})\wedge d\psi\wedge e^{-F_\nu^{\rm D5}}.
\label{rrvvx}
\end{equation}
If we turn on the $\U(1)_R$ gauge field, $d\psi$ is replaced by $d\psi+2V_R$,
and the $\U(1)_R$ gauge transformation changes it by
$2d\lambda_R$.
Therefore, the $\U(1)_R$ gauge transformation of (\ref{rrvvx}) is
\begin{equation}
\delta C^{\rm mag}|_{\theta=0}=d\lambda_R\wedge \sum_\nu \frac{1}{\pi}[[\phi-\phi_\nu-\pi]]
\delta({\bm\nu})\wedge e^{-F_\nu^{\rm D5}/2\pi}.
\label{drrvv}
\end{equation}
The anomaly flow associated with this gauge transformation is
\begin{eqnarray}
\delta S_{\rm D5}
&=&\frac{1}{2}\sum_\mu\int_{\partial{\rm D5}}\delta C|_\mu\wedge e^{F_\mu^{\rm D5}/2\pi}
\nonumber\\
&=&\frac{1}{16\pi^2}\sum_{\mu,\nu}\frac{1}{\pi}[[\phi_\mu-\phi_\nu-\pi]]
\langle{\bm\mu},{\bm\nu}\rangle\nonumber\\
&&\quad
\int \lambda_R(F_\mu^{\rm D5}-F_\nu^{\rm D5})
\wedge(F_\mu^{\rm D5}-F_\nu^{\rm D5})
\end{eqnarray}
This coincides the same with the anomaly computed in the gauge theory.

Up to now we have assumed that the integration constant $c$ in
(\ref{eq113}) vanishes.
Before ending this section,
let us discuss the $c$ independence of the
anomaly flow.
The extra term arising in the gauge transformation
of the RR potential due to the non-vanishing integration constant $c$
is
\begin{equation}
\delta C
=\frac{1}{\pi}d\lambda_R\wedge c.
\label{drrvvc}
\end{equation}
With this gauge transformation,
we obtain the following extra contribution to the
anomaly flow:
\begin{eqnarray}
\delta S_{\rm D5}
&=&\frac{1}{2\pi}\sum_\mu\int_{\partial{\rm D5}} \lambda_R c\wedge e^{F_\mu^{\rm D5}/2\pi}
\nonumber\\
&=&\frac{1}{16\pi^3}\sum_\mu\int_{{\bf R}^4\times T} \lambda_R c_1\wedge\delta({\bm\mu})\wedge F_\mu^{\rm D5}\wedge F_\mu^{\rm D5}
\nonumber\\
&=&\frac{N}{2\pi}\int_{{\bf R}^4\times T} \lambda_R c_1\wedge X.
\label{c1dep}
\end{eqnarray}
Between the first and second lines in (\ref{c1dep}) we used
\begin{equation}
\sum_\mu\delta({\bm\mu})=
\sum_\mu\delta({\bm\mu})\wedge F_\mu^{\rm D5}=0.
\end{equation}
Due to this relation
(\ref{c1dep}) includes only the one-form part $c_1$ of the formal sum
$c\equiv c_1+c_3+c_5$.

The $c_1$ dependent contribution
(\ref{c1dep}) is canceled by taking account of the gauge transformation
of the bulk action%
\footnote{The coefficient of this Chern-Simons term is half
of the coefficient of the Chern-Simons term used for the purpose of
obtaining the equations of motion for the gauge fields.
This is because we are computing a kind of ``self-energy'' of the system.
This may be related to the subtlety associated with the self-dual field
we often meet when we consider the action
of type IIB supergravity.
}
\begin{equation}
S_{\rm bulk}=\frac{1}{8\pi^2}\int C_2\wedge H_3\wedge G_5.
\end{equation}
Indeed, if we
perform $c_1$ dependent part of the $\U(1)_R$
gauge transformation
\begin{equation}
\delta C_2=\frac{1}{\pi}d\lambda_R\wedge c_1
\label{cdepend}
\end{equation}
and substitute the term
\begin{equation}
G_5=-2\pi X,
\end{equation}
we obtain
\begin{eqnarray}
\delta S_{\rm bulk}
&=&-\frac{1}{4\pi^2}\int_{{\bf R}^4\times T\times{\bf B}_4}d\lambda_R\wedge c_1\wedge H_3\wedge X
\nonumber\\
&=&-\frac{1}{4\pi^2}\int_{{\bf R}^4\times T\times{\bf S}^3}\lambda_R c_1\wedge H_3\wedge X.
\label{delsbulk}
\end{eqnarray}
In the first line of (\ref{delsbulk}),
the integration region is the direct product of
${\bf R}^4$ along $0123$, $T$, and four dimensional solid ball in $4689$ space.
To obtain the second line we used Stokes' theorem and
${\bf S}^3$ in the integration region is the boundary of ${\bf B}_4$.
By using $\oint_{{\bf S}^3} H_3=2\pi N$ we obtain
\begin{equation}
\delta S_{\rm bulk}
=-\frac{N}{2\pi}\int_{{\bf R}^4\times T}\lambda_R c_1\wedge X.
\end{equation}
This precisely cancels the $c$ dependent term (\ref{c1dep}).

\section{Conclusions and Discussions}
In this paper we studied the brane realization of
global symmetries and 't~Hooft anomalies associated with them.

We showed that $\U(1)_B$ symmetries
can be realized as linear combinations of
the $\U(1)$ gauge symmetries
on the fivebranes,
while the $\U(1)_M$ symmetries
are combinations of gauge symmetry of the NS-NS $2$-form field in the bulk
and the $\U(1)$ gauge symmetries on the NS5-branes.
The mixing ambiguity for $\U(1)_M$
can be interpreted as the gauge ambiguity
associated with the $B$-field gauge transformation.

We identified $\U(1)_R$ symmetry
with the rotation of the system on the $8$-$9$ plane.
With this identification,
it seems that there is a preferred $R$-charge assignment
of the bi-fundamental fields.
Namely, the angles among external legs seems to determine the R-charges
uniquely.
It is not clear how we should interpret this specific $\U(1)_R$ symmetries
on the gauge theory side.
One possibility is that it may be related to the $\U(1)_R$
symmetry in the superconformal algebra.
We cannot, however, simply identify these two.
One reason is as follows.
The charges for superconformal $\U(1)_R$ is determined
with the help of so-called $a$-maximization procedure,
and the charges obtained by this procedure are always quadratic rational numbers.
Namely, they are represented as $a+b\sqrt{m}$ with rationals $a$ and $b$
and an integer $m$.
On the other hand, in \S\S\ref{rsym.sec},
R-charges are given as angles among external legs in web diagrams.
It is unlikely that these two sets of quantities,
a set of quadratic rationals and a set of angles, coincide.
Another reason is that the angles among the external legs depend
on the modulus of the torus along 57 directions,
while the $a$-maximizing R-charge assignment does not seem to have
such a degree of freedom.
We may be able to understand the disagreement between the two sets of R-charges
as follows.
In this paper we only used the asymptotic shape
of the branes to determine the charges.
This seems to give only the information of ultra-violet region.
On the other hand, the $a$-maximizing symmetry is in general realized only
in the infra-red limit.
Thus, it is natural that the two sets of charges are different.
In order to investigate low-energy dynamics of gauge theories,
including the charges of operators for the superconformal R-symmetry,
it would be important to know the precise shape of the brane
in the central region of the brane configuration.

In \S\ref{anomaly.sec}, we showed that two kinds of
non-vanishing 't~Hooft anomalies,
$\U(1)_M\U(1)_B^2$ and $\U(1)_R\U(1)_B^2$ anomalies,
are obtained as the variations of the classical action of the branes.
Natural question arises here is that how we can get other anomalies.
Especially, $\U(1)_R^3$ and $\U(1)_R$,
which are used in the $a$-maximization,
are very important if we
discuss low-energy dynamics of gauge theories.
Because R-symmetry is identified with the
spacial rotation of the brane system,
the corresponding gauge field enters in
the metric and the spin connection.
Therefore, in order to obtain
$\U(1)_R^3$ and $\U(1)_R$ anomalies, we need to consider
the brane configuration in curved backgrounds,
and take account of the
higher derivative terms including the
curvature of the background spacetime.
This (and the problem about R-charges we mentioned above)
makes the analysis more difficult
than what we have done in this paper.

\section*{Acknowledgements}
I would like to thank
A.~Kato, F.~Koyama,
Y.~Nakayama, R.~Suzuki and M.~Yamazaki
for valuable comments and discussions.
I would also like to thank the Yukawa Institute for
Theoretical Physics at Kyoto University.
Discussions during the YITP workshop YITP-W-06-11
on ``String Theory and Quantum Field Theory''
were useful to complete this work.
This work is supported in part by
a Grant-in-Aid for Scientific Research
(\#17540238)
from the Japan Ministry of Education, Culture, Sports,
Science and Technology.

\appendix
\section{The action of $(p,q)$ fivebrane junctions}\label{junc.sec}
In this appendix, we determine the boundary condition imposed on
gauge fields on fivebranes at junctions
by requiring the gauge invariance of the action.
We first fix our conventions.

In this appendix, we use different normalization for gauge fields
and actions.
Any gauge field $A$ (a gauge field on a brane or a bulk gauge field)
in this appendix is related to the
field $A'$ in the other sections by rescaling $A=A'/(2\pi)$.
An action $S$ in this appendix is similarly rescaled by $S=S'/(2\pi)$
where $S'$ is the action in the usual normalization.
This rescaling removes $2\pi$ from the following equations.

The electric-magnetic duality relations for
the RR field strengths and the NS-NS field strengths in type IIB supergravity
are
\begin{equation}
G_5=-*G_5,\quad
G_7=*G_3,\quad
G_9=-*G_1,\quad
H_7=-e^{-2\phi}*H_3.
\end{equation}
The first one represents the (anti-)self-duality of
the RR $5$-form field strength.

Let $G=G_1+G_3+G_5+G_7+G_9$ be the formal sum of RR
field strengths.
The Bianchi identities and equations of motion in the string frame are
\begin{equation}
dG=H_3\wedge G
\end{equation}
\begin{equation}
dH_3=0.
\end{equation}
\begin{equation}
dH_7=G_3\wedge G_5-G_1\wedge G_7
\label{h7bi}
\end{equation}
We define the NS-NS potential $B_2$ by
\begin{equation}
H_3=dB_2.
\label{h3db2}
\end{equation}
The RR potentials are defined by
\begin{equation}
G=dC-H_3\wedge C
=e^{B_2}\wedge d(e^{-B_2}\wedge C),
\label{Gflux}
\end{equation}
where $C$ is the formal sum
$C=C_0+C_2+C_4+C_6+C_8$.

The field strengths $G$ and $H_3$ above are invariant under
the $B$-field gauge transformations
\begin{equation}
\delta B_2=d\Lambda_1^B,
\label{nstr}
\end{equation}
and the RR gauge transformations
\begin{equation}
\delta C=e^{B_2}\wedge d\Lambda^C,
\label{rrtr}
\end{equation}
where $\Lambda^C$ is the formal sum
of the parameters $\Lambda^C=\Lambda_1^C+\Lambda_3^C+\Lambda_5^C+\Lambda_7^C$.
As we see below, we can determine the transformation law for
NS-NS magnetic potential $B_6$ in such a way that $H_7$ is also gauge invariant.

With this conventions for the gauge fields,
the D5-brane Chern-Simons action is given by
\begin{equation}
S_{(1,0)}=\int C\wedge e^{{\cal F}^{(1,0)}},
\label{105action}
\end{equation}
where the field strength ${\cal F}^{(1,0)}$
is given by ${\cal F}^{(1,0)}=dV^{(1,0)}-B_2$.
The gauge field $V^{(1,0)}$ is transformed under the $B$-field gauge transformation (\ref{nstr}) by
\begin{equation}
\delta V^{(1,0)}=\Lambda^B_1.
\label{da10}
\end{equation}

Before we determine the gauge invariant action
for fivebrane junctions,
we discuss the $(p,q)$ fivebrane actions.
They are obtained from the D5-brane action by $\SL(2,{\bf Z})$ transformations.
We should note that some of gauge fields we defined above are
not $\SL(2,{\bf Z})$ covariant, and we should change the basis before performing
$\SL(2,{\bf Z})$ transformations.
The $\SL(2,{\bf Z})$ invariant metric,
the Einstein metric, is
obtained by the Weyl rescaling
\begin{equation}
\wt g_{\mu\nu}=e^{-\phi/2}g_{\mu\nu}.
\end{equation}
Similarly, we define $\SL(2,{\bf Z})$ covariant RR $3$-form $\wt G_3$
and NS-NS $7$ form $\wt H_7$ by
\begin{equation}
\wt G_3=G_3+C_0H_3,\quad
\wt H_7=H_7+C_0G_7.
\end{equation}
The $3$-form field strengths
$(H_3,\wt G_3)$ and
the $7$-form field strengths $(G_7,\wt H_7)$
are $\SL(2,{\bf Z})$ doublets transformed by
\begin{equation}
\left(\begin{array}{c}
G_7 \\
\wt H_7
\end{array}\right)
\rightarrow
\left(\begin{array}{cc}
p & q \\
r & s
\end{array}\right)
\left(\begin{array}{c}
G_7 \\
\wt H_7
\end{array}\right),\quad
\left(\begin{array}{c}
H_3 \\
\wt G_3
\end{array}\right)
\rightarrow
\left(\begin{array}{cc}
p & q \\
r & s
\end{array}\right)
\left(\begin{array}{c}
H_3 \\
\wt G_3
\end{array}\right).
\label{sl2z}
\end{equation}
By this $\SL(2,Z)$ transformation,
the complex field $\tau=C_0+ie^{-\phi}$ is transformed by
\begin{equation}
\tau\rightarrow
\tau'=\frac{s\tau+r}{q\tau+p}.
\label{tautr}
\end{equation}
The self-dual field strength $G_5$ is $\SL(2,{\bf Z})$ invariant as it is.

For the potentials,
we introduce the following ones:
\begin{equation}
\wt C_4=C_4-\frac{1}{2}B_2\wedge C_2,\quad
\wt C_6=C_6-\frac{1}{6}B_2\wedge B_2\wedge C_2.
\end{equation}
$\wt C_4$ is $\SL(2,{\bf Z})$ invariant and
$\wt C_6$ forms $\SL(2,{\bf Z})$ doublet together with the NS-NS magnetic potential
$B_6$ defined by
\begin{equation}
\wt H_7=dB_6-\wt C_4\wedge\wt G_3-\frac{1}{6}C_2\wedge(C_2\wedge dB_2-B_2\wedge dC_2).
\end{equation}
The gauge transformation law for $B_6$ which keeps the field strength $\wt H_7$
invariant is
\begin{equation}
\delta B_6
=d\Lambda_5^B
+C_2\wedge d\Lambda_3^C
+\frac{1}{3}C_2\wedge(B_2\wedge d\Lambda_1^C-C_2\wedge d\Lambda_1^B).
\end{equation}
Under the $\SL(2,{\bf Z})$ transformation
(\ref{sl2z}),
the two-form and six-form potentials are transformed as
\begin{equation}
\left(\begin{array}{c}
\wt C_6 \\
B_6
\end{array}\right)
\rightarrow
\left(\begin{array}{cc}
p & q \\
r & s
\end{array}\right)
\left(\begin{array}{c}
\wt C_6 \\
B_6
\end{array}\right),\quad
\left(\begin{array}{c}
B_2 \\
C_2
\end{array}\right)
\rightarrow
\left(\begin{array}{cc}
p & q \\
r & s
\end{array}\right)
\left(\begin{array}{c}
B_2 \\
C_2
\end{array}\right).
\label{sl2zpot}
\end{equation}

The $(p,q)$ fivebrane action is obtained from the D5-brane action
(\ref{105action}) by the replacement (\ref{sl2zpot}).
For example, the minimal coupling term $\int\wt C_6$ in the
D5-brane action is transformed as $\int(p\wt C_6+qB_6)$.
This term implies that the fivebrane possesses the
fivebrane charge $(p,q)$.
We should note that there is an ambiguity for the choice of the
$\SL(2,{\bf Z})$ element.
The shift $(r,s)\rightarrow(r,s)+n(p,q)$ with arbitrary $n$
does not change the $(p,q)$ charge, and thus we should have
the same fivebrane action regardless of the choice of the $\SL(2,{\bf Z})$ element.
We will return to this point later.

The charge $p$ and $q$ obtained by the $\SL(2,{\bf Z})$ transformation
are always co-prime.
For the purpose of treating coincident fivebranes,
it is convenient to relax this condition,
and extend the duality group $\SL(2,{\bf Z})$ to $\SL(2,{\bf Q})$.
If charges $p$ and $q$ is not co-prime, and $n\equiv {\rm GCD}(p,q)\neq1$,
the brane is regarded as a stack of $n$ elementary branes.
In this case $\U(n)$ gauge field $\wh A^{(p,q)}$ lives on the worldvolume.
We assume that the $\SU(n)$ part of $\wh A^{(p,q)}$ vanishes and define the
diagonal part $A^{(p,q)}$ by
\begin{equation}
A^{(p,q)}=\tr\wh A^{(p,q)}.
\end{equation}
The action of the stack of branes is obtained by $\SL(2,{\bf Q})$ transformation
from the D5-brane action.
For example, the action of
$(n,0)$ fivebrane, a stack of $n$ D5-branes,
is
\begin{eqnarray}
S_{(n,0)}
&=&\int C\wedge\tr e^{\wh{\cal F}^{(n,0)}}
\nonumber\\
&=&\int\Big(nC_6
+C_4\wedge{\cal F}^{(n,0)}
+\frac{1}{2n}C_2\wedge{\cal F}^{(n,0)}\wedge{\cal F}^{(n,0)}
\nonumber\\&&\hspace{3em}
+\frac{1}{6n^2}C_0\wedge{\cal F}^{(n,0)}\wedge{\cal F}^{(n,0)}\wedge{\cal F}^{(n,0)}\Big),
\end{eqnarray}
where
\begin{equation}
\wh{\cal F}^{(n,0)}=d\wh V^{(n,0)}-{\bf 1}_nB_2,\quad
{\cal F}^{(n,0)}=
\tr\wh {\cal F}^{(n,0)}
=dV^{(n,0)}-nB_2.
\end{equation}
The above action is obtained as the $\SL(2,{\bf Q})$ transformation
of D5-brane action with the element
\begin{equation}
\left(\begin{array}{cc}
n \\
& 1/n
\end{array}
\right)\in \SL(2,{\bf Q}).
\end{equation}
In what follows, we do not require the co-primeness of the charges
$p$ and $q$.

We here do not give the full $(p,q)$ fivebrane action
explicitly because even though it can be obtained straightforwardly
by the $\SL(2,{\bf Z})$ transformation,
the expression is not simple due to the basis change.
Fortunately, we can show that the gauge transformation parameters
$\Lambda^C$, $\Lambda^B_1$ and $\Lambda^B_5$ are $\SL(2,{\bf Z})$ covariant
without any basis change.
$(\Lambda_1^B,\Lambda_1^C)$ and $(\Lambda_5^C,\Lambda_5^C)$ are
$\SL(2,{\bf Z})$ doublets  transformed as
\begin{equation}
\left(\begin{array}{c}
\Lambda^C_5 \\
\Lambda^B_5
\end{array}\right)
\rightarrow
\left(\begin{array}{cc}
p & q \\
r & s
\end{array}\right)
\left(\begin{array}{c}
\Lambda^C_5 \\
\Lambda^B_5
\end{array}\right),\quad
\left(\begin{array}{c}
\Lambda^B_1 \\
\Lambda^C_1
\end{array}\right)
\rightarrow
\left(\begin{array}{cc}
p & q \\
r & s
\end{array}\right)
\left(\begin{array}{c}
\Lambda^B_1 \\
\Lambda^C_1
\end{array}\right),
\label{sl2zpara}
\end{equation}
and $\Lambda_3^C$ is $\SL(2,{\bf Z})$ invariant.
By this reason, we only show the
gauge transformations for the $(p,q)$ fivebrane actions
obtained by the $\SL(2,{\bf Z})$ transformation of the
transformation of the D5-brane action.

The field strength ${\cal F}^{(1,0)}$ in the D5-brane action
(\ref{105action}) is invariant
under the $B_2$ gauge transformation,
and so is the action $S_{(1,0)}$.
With respect to the RR gauge transformation (\ref{rrtr}),
the D5-brane action is gauge invariant up to the
following boundary variation:
\begin{eqnarray}
\delta S_{(1,0)}
&=&\int_{\partial D5} \Lambda^C\wedge e^{F_2^{(1,0)}}
\nonumber\\
&=&\int_{\partial D5}\left(
\Lambda_5^C
+\Lambda_3^C\wedge F_2^{(1,0)}
+\frac{1}{2}\Lambda_1^C\wedge F_2^{(1,0)}\wedge F_2^{(1,0)}\right)
\label{delsd5}
\end{eqnarray}
From this variation, we can easily obtain the
variation of the $(p,q)$ fivebrane action.
We replace the potentials and transformation parameters
according to (\ref{sl2zpot}) and (\ref{sl2zpara}),
and $V^{(1,0)}$ is renamed $V^{(p,q)}$.
As the result, we obtain
\begin{equation}
\delta S_{(p,q)}
=\int_{\partial(p,q)}\left(
(p\Lambda_5^C+q\Lambda_5^B)
+\Lambda_3^C\wedge F_2^{(p,q)}
+\frac{1}{2}(r\Lambda_1^B+s\Lambda_1^C)\wedge F_2^{(p,q)}\wedge F_2^{(p,q)}\right)
\end{equation}
From the gauge transformation
of the gauge field (\ref{da10}), we obtain
\begin{equation}
\delta V^{(p,q)}=p\Lambda_1^B+q\Lambda_1^C.
\label{dapq}
\end{equation}

Let us consider a $3$-fivebrane junction
consisting of three fivebranes with charges $(p_i,q_i)$ ($i=1,2,3$).
The variation arising on the junction is
\begin{eqnarray}
\sum_{i=1}^3\delta S_{(p_i,q_i)}
&=&\int_J
\Big(
(p_1+p_2+p_3)\Lambda_5^C+(q_1+q_2+q_3)\Lambda_5^B
\nonumber\\
&&\hspace{5em}+\Lambda_3^C\wedge
(F_2^{(1)}+F_2^{(2)}+F_2^{(3)})
\nonumber\\
&&\hspace{5em}
+\frac{1}{2}\sum_{i=1}^3(r_i\Lambda_1^B+s_i\Lambda_1^C)
\wedge F_2^{(i)}\wedge F_2^{(i)}\Big),
\label{subdels}
\end{eqnarray}
where $\int_J$ means the integration over the junction,
and $F^{(i)}\equiv F^{(p_i,q_i)}$.

For the cancellation of the first and second lines,
we should impose
the charge conservation condition
\begin{equation}
p_1+p_2+p_3=q_1+q_2+q_3=0,
\end{equation}
and the boundary condition for the gauge fields
\begin{equation}
F_2^{(1)}
+F_2^{(2)}
+F_2^{(3)}=0.
\label{fffvc}
\end{equation}
The condition (\ref{fffvc}) for the field strengths
is satisfied if
\begin{equation}
V^{(1)}+V^{(2)}+V^{(3)}=0.
\end{equation}

The variation in the third line in (\ref{subdels})
is canceled by introducing
the following action:
\begin{eqnarray}
S_J&=&\frac{1}{6\Delta}\int_J(
c_{111} V^{(1)}\wedge F^{(1)}\wedge F^{(1)}
+c_{112} V^{(1)}\wedge F^{(1)}\wedge F^{(2)}
\nonumber\\
&&\hspace{3em}
+c_{122} V^{(1)}\wedge F^{(2)}\wedge F^{(2)}
+c_{222} V^{(2)}\wedge F^{(2)}\wedge F^{(2)}
),
\label{sjunc}
\end{eqnarray}
where $\Delta$ and the coefficients $c_{ijk}$ are
defined by
\begin{equation}
\Delta
=p_1q_2-p_2q_1
=p_2q_3-p_3q_2
=p_3q_1-p_1q_3,
\end{equation}
and
\begin{eqnarray}
c_{111}&=&p_2(s_1+s_3)-q_2(r_1+r_3),\nonumber\\
c_{112}&=&3(p_2s_3-q_2r_3),\nonumber\\
c_{122}&=&-3(p_1s_3-q_1r_3),\nonumber\\
c_{222}&=&-p_1(s_2+s_3)+q_1(r_2+r_3).
\end{eqnarray}
As we mentioned above, there is the ambiguity for the choice
of $\SL(2,{\bf Z})$ element when we obtain the $(p,q)$ fivebrane actions
form the D5-brane action.
Let $p_i$, $q_i$, $r_i$ and $s_i$ be the components of the $\SL(2,{\bf Z})$
matrices used to obtain $S_{(p_i,q_i)}$.
The action (\ref{sjunc}) includes $r_i$ and $s_i$,
and depend on the choice of these parameters.
By the shift
\begin{equation}
(r_i,s_i)\rightarrow
(r_i,s_i)+n_i(r_i,s_i),
\label{rsshifts}
\end{equation}
the action
(\ref{sjunc}) is shifts by
\begin{equation}
S_J
\rightarrow S_J
-\frac{1}{6}\int_J\sum_{i=1}^3
n_i V^{(i)}\wedge F^{(i)}\wedge F^{(i)}.
\label{sjshift}
\end{equation}
As we mentioned above,
the bulk part $S_{(p_i,q_i)}$ of the fivebrane action
also shifted by (\ref{rsshifts}).
It is given by
\begin{equation}
S_{(p_i,q_i)}
\rightarrow S_{(p_i,q_i)}+\frac{n_i}{6}\int_{(p_i,q_i)} F_i\wedge F_i\wedge F_i.
\label{blkshift}
\end{equation}
This is precisely canceled
by the shift
(\ref{sjshift}),
and the total action $\sum S_{(p_i,q_i)}+S_J$ is
invariant under (\ref{rsshifts}).

A convenient choice of $(r_i,s_i)$ is
\begin{equation}
(r_1,s_1)=\frac{1}{2\Delta}(p_2-p_3,q_2-q_3).
\end{equation}
$(r_2,s_2)$ and $(r_3,s_3)$ are given by similar equations with
cyclically permuted indices.
In this case, the junction action becomes simple and cyclically symmetric
\begin{eqnarray}
S_J
&=&-\frac{1}{12\Delta}\int\Big[
(V^{(1)}-V^{(2)})\wedge F^{(3)}\wedge F^{(3)}
\nonumber\\&&\hspace{4em}
+(V^{(2)}-V^{(3)})\wedge F^{(1)}\wedge F^{(1)}
\nonumber\\&&\hspace{5em}
+(V^{(3)}-V^{(1)})\wedge F^{(2)}\wedge F^{(2)}
\Big].
\label{symsj}
\end{eqnarray}
The terms in this action look like electric coupling between
the string charges induced by the fluxes on the fivebranes
and the gauge potentials on other branes.

In order to clarify the meaning of this action
let us consider a simple case,
the $(1,m_1)$-$(-1,-m_2)$-$(n,0)$ junction.
For the charge conservation,
$m_2-m_1=n$ must hold.
Let us regard two of the three semi-infinite fivebranes,
the $(1,m_1)$ fivebrane and the $(-1,-m_2)$ fivebrane,
as two parts of one infinite brane separated by the
junction.
This is natural especially when the coupling constant $g_{\rm str}$
is small because in this case the tensions of the $(1,m_1)$ and
$(-1,-m_2)$ branes
are almost the same with the tension of NS5-brane
and are much larger than the tension of $(n,0)$ brane.
We can treat two large-tension branes as one NS5-brane
with zero-form flux carrying the D5-brane charge
$m_1$ and $m_2$.
The gauge fields $V^{(m_1,1)}$ and $V^{(-m_2,-1)}$
can be identified with the restriction of the gauge field $A_{\rm NS5}$
in two regions.
\begin{equation}
V_{\rm NS5}=
\left\{\begin{array}{ll}
V^{(m_1,1)} & \mbox{in $(m_1,1)$ fivebrane}\\
-V^{(-m_2,-1)} & \mbox{in $(-m_2,-1)$ fivebrane}
\end{array}\right.
\end{equation}
In this case (\ref{symsj}) becomes
\begin{eqnarray}
S_J
&=&
-\frac{1}{4n}\int
(V^{(1)}-V^{(2)})\wedge F^{(3)}\wedge F^{(3)}
\nonumber\\&&
+\frac{1}{12n}\int(
V^{(1)}\wedge F^{(1)}\wedge F^{(1)}
-V^{(2)}\wedge F^{(2)}\wedge F^{(2)}
).
\end{eqnarray}
The two terms in the second line can be rewritten as bulk terms,
and only the first term is essentially the boundary term.
If we define the gauge field in NS5-brane on the junction
as the average of two region by
\begin{equation}
\ol V=\frac{1}{2}(V^{(1)}-V^{(2)}),
\end{equation}
the boundary interaction becomes
\begin{equation}
S
=-\frac{1}{2n}\int_J
\ol V\wedge F^{(3)}\wedge F^{(3)}
=-\frac{1}{2}\int_J
\ol V\wedge\tr(\wh F^{(3)}\wedge \wh F^{(3)}).
\end{equation}
This represents the coupling of the D-string current on the D5-brane
and the gauge fields on NS5-branes.

\section{Solution to the Maxwell equation}\label{maxwell.sec}
In this appendix, we give the solution to the Maxwell equation
with a conserved current localized on a two-dimensional plane.
The solution is used in \S\S\ref{rbb.ssec}
to compute the anomaly flow.

Let $j_3$ and $F=dA$ be the conserved current three-form
and a $\U(1)$ gauge field
coupling to the current defined in the $4689$ space.
They satisfy the Maxwell equation
\begin{equation}
dF=j_3,\quad
d*F_2=0.
\label{maxwell}
\end{equation}
The conservation of $j_3$ means that $dj_3=0$, and
we can define the ``magnetization'' two-form $M_2$ by $j_3=dM_2$.
In the case that $j_3$ has its support in the $46$-plane,
we can take $M_2$ as
\begin{equation}
M_2=m(x_4,x_6)\delta(x_8)dx_8\wedge \delta(x_9)dx_9
\end{equation}
with a function $m$ defined on the $46$ plane.

To solve the equations in (\ref{maxwell}),
we start from the following ansatz:
\begin{eqnarray}
F=\frac{1}{2\pi}df\wedge d\psi,
\label{fanz}
\end{eqnarray}
where we introduced the polar coordinates $(r,\psi)$ by $x^8+ix^9=re^{i\psi}$,
and $f$ is a function of $x^4$, $x^6$ and $r$.
Because
$d(d\psi)=2\pi\delta(x^8)dx^8\wedge \delta(x^9)dx^9$,
the exterior derivative of this field strength
has its support at $r=0$:
\begin{equation}
dF
=df(x^4,x^6,r=0)\wedge\delta(x^8)\delta(x^9)dx^8\wedge dx^9.
\end{equation}
Thus (\ref{fanz}) satisfies $dF=j_3$ provided that the
function $f$ satisfies the boundary condition
\begin{equation}
f(x^4,x^6,r=0)=m(x^4,x^6).
\label{fbdr}
\end{equation}
The other equation $d*F=0$ is satisfied if the function $f$
is a solution of
\begin{equation}
\left(
\partial_4^2+\partial_6^2
+r\partial_r\frac{1}{r}\partial_r
\right)f(x^4,x^6,r)=0.
\label{ddf}
\end{equation}
The solution for
(\ref{fbdr}) and (\ref{ddf}) is obtained by using the Green function as
\begin{equation}
f(x^4,x^6,r)
=\frac{1}{\pi}\int dx\int dy\frac{r^2m(x,y)}{[(x_4-x)^2+(x_6-y)^2+r^2]^2}.
\end{equation}
We can give the potential for the field strength $F$ in (\ref{fanz})
as
\begin{equation}
A=\frac{1}{2\pi}f d\psi
\end{equation}
On the $46$-plane,
the gauge potential is proportional to the magnetization.
\begin{equation}
A|_{r=0}=\frac{1}{2\pi}m(x,y)d\psi,
\end{equation}

In \S\S\ref{rbb.ssec} we need the solution to (\ref{maxwell})
with a current in the form
\begin{equation}
j_3=\int_0^{2\pi}d\phi \rho(\phi)\delta_3(L_\phi).
\end{equation}
For this current, the magnetization is given by
\begin{equation}
m(\phi)=\frac{1}{2\pi}\int{d\phi'}\rho(\phi')[[\phi-\phi'-\pi]]+c,
\end{equation}
where $c$ is an integration constant.


\end{document}